\documentclass[showpacs, pra,onecolumn,preprintnumbers ,amsmath, amssymb, superscriptaddress, aps]{revtex4-2}
\usepackage{color}
\usepackage{amsmath,amssymb}
\usepackage{pifont}% http://ctan.org/pkg/pifont
\usepackage{amssymb}  % More symbols
\usepackage{bbold}
\usepackage{float}
\usepackage{subfloat}
\usepackage[squaren]{SIunits}

\usepackage[caption=false]{subfig}
\usepackage{tikz}
\usepackage{makecell}
\usepackage{subfig}
\usepackage{pifont}   % Ding symbols
\usepackage{graphicx} % Include figure files
\graphicspath{{Figuress/}}
\usepackage{dcolumn}  % Align table columns on decimal point
\usepackage{bm}       % bold math
\usepackage{multirow} % Table functions
\usepackage{placeins}% For making floats not move around everywhere
\usepackage[colorlinks]{hyperref}
\usepackage{mathtools}
\usepackage{appendix}
\usepackage{amsmath}
\usepackage{appendix}

\newcommand\summaryname{Abstract}
    {\small\begin{center}%
    \bfseries{\summaryname} \end{center}}

\usepackage{fancyhdr}
\begin{document}
\title{\textbf{Einstein-Podolsky-Rosen Steering in Three Coupled Harmonic Oscillators}}
   \author{Ayoub Ghaba}
       \email{ayoub.ghaba-etu@etu.univh2c.ma}
 \affiliation{Laboratory of Mathematics and Physical Sciences Applied to Engineering Sciences, Faculty of Sciences and Techniques, Hassan II University, Mohammedia, Morocco}
  \author{Radouan Hab Arrih}
        \email{habarrih46@gmail.com}
 \affiliation{Laboratory of R\&D in Engineering Sciences, Faculty of Sciences and Techniques Al-Hoceima,
Abdelmalek Essaadi University, Tetouan, Morocco.}
\author{Elhoussine Atmani}
\affiliation{Laboratory of Mathematics and Physical Sciences Applied to Engineering Sciences, Faculty of Sciences and Techniques, Hassan II University, Mohammedia, Morocco}
\affiliation{Laboratory of R\&D in Engineering Sciences, Faculty of Sciences and Techniques Al-Hoceima,
Abdelmalek Essaadi University, Tetouan, Morocco.}\author{Abdallah Slaoui}\email{abdallah.slaoui@um5s.net.ma}\affiliation{LPHE-MS, Faculty of Sciences, Mohammed V University in Rabat, Rabat, Morocco}\affiliation{CPM, Faculty of Sciences, Mohammed V University in Rabat, Rabat, Morocco.}
\date{\today}
\maketitle
 	\begin{center}
 	\textbf{Abstract}
 \end{center}
%\begin{abstract}
Quantum steering is one of the most intriguing phenomena in quantum mechanics and is essential for understanding correlations in multi-body systems. Despite its importance, analytical results for coupled three-body oscillators remain scarce. In this work, we investigate this phenomenon through a geometrical diagonalization approach, which reduces the degrees of freedom associated with the system's steering properties. Specifically, we derive analytical expressions for quantum steering in all possible directions using the Wigner function framework, as it provides a complete description of the system's quantum state. Our results indicate that excitations significantly enhance quantum steering across the system; this stands in contrast to the ground state $(0,0,0)$, which exhibits no steerable correlations. Furthermore, both the directionality and topology of these correlations are governed by the spatial distribution of the excitations rather than their magnitude. We also observe symmetric steering behavior between oscillators $x$, $y$, and $z$ under equivalent excitation conditions, which can be formalized as $S^{(n,m,l)}_{x\to z}(\theta)=S^{(n,m,l)}_{x\to y}(-\theta),\quad S^{(n,m,l)}_{z\to x}(\theta)=S^{(n,m,l)}_{y\to x}(-\theta)$, and $S^{(n,m,l)}_{y\to z}(\theta)=S^{(n,m,l)}_{z\to y}(-\theta)$. Therefore, we elucidate how excitation levels and mixing angles generate and enhance steering in three coupled harmonic oscillators.\\

  	{\sc Keywords:} Quantum steering, three coupled oscillators, Wigner function, quantum fluctuations, geometrical diagonalization
%\end{abstract}

\section{Introduction}
One of the most fascinating phenomena in quantum mechanics is quantum steering, which has attracted considerable interest in both quantum mechanics and quantum information theory. In fact, Schrödinger was the first to propose \cite{Introduction1} that one part of a composite system could steer the state of the other part toward a specific position or momentum state by choosing an appropriate measurement \cite{intro1}. Earlier, in 1935, Einstein, Podolsky, and Rosen (EPR) presented their famous argument against the completeness of quantum mechanics \cite{intro1,intro2}. In this reasoning, a two-particle state is assumed, in which one part can measure either the position or the momentum of the other part. The correlations of the state allow the outcomes of these measurements on the other part to be predicted if the same measurement is performed there. Schrödinger’s idea emerged in response to this. The distinction between this nonlocal influence and classical correlations, or even entanglement, lies in the fact that it cannot be explained by classical theories based on local hidden variables. The formalism of quantum steering was later developed and expanded by Wiseman, Jones, and Doherty \cite{intro3}, who placed it between quantum entanglement and Bell nonlocality.\par

Quantum steering can occur even when systems are not maximally entangled, distinguishing it from entanglement which requires strong correlations between distant systems. This characteristic makes quantum steering a unique form of quantum correlation that exhibits nonlocal effects, yet it remains less stringent than Bell nonlocality, which arises from violations of Bell's inequalities. Quantum steering is demonstrated when no local hidden state model can account for the measurement outcomes of a subsystem, thereby proving that the correlations are inherently quantum rather than the result of classical interactions. Broadly defined, quantum steering refers to a bipartite scenario in which one party can influence (or steer) the state of a distant party through local measurements \cite{intro4}. Reid \cite{intro5,intro6,intro7,intro8} demonstrated this experimentally for the first time in a continuous-variable (CV) system via non-degenerate parametric amplification in the 1980s. Such an approach has since been successfully used to study EPR steering in a wide range of continuous-variable systems \cite{intro7,intro12}. Our current understanding of quantum steering detection and distribution has advanced considerably \cite{intro13,intro14,intro15}. Steering is an important quantum resource in quantum information and computation \cite{intro16}, and it plays a crucial role in quantum teleportation \cite{Ikken2025,Slaoui2024}, as well as in secure communication \cite{intro17} and quantum network security \cite{intro18}.\par

The harmonic oscillator formalism serves as a fundamental tool in physics, appearing across a diverse array of theoretical frameworks. It is notably employed in describing bimodal squeezed states of light \cite{Han1990,Shanta1996}, Lee's model in quantum field theory \cite{Schweber2005}, the covariant harmonic oscillator model in parton theory \cite{kim1989}, and the Bogoliubov transformation in superconductivity \cite{Fetter2012}, as well as various approaches to molecular physics \cite{Iachello1991}. Furthermore, several models rely on unobservable degrees of freedom, such as those utilized in two-mode squeezed states \cite{Yurke1987}, hadronic temperature models \cite{Han1989}, and in the Barnett–Phoenix formulation of information science \cite{Barnett1991}. The behavior of coupled harmonic oscillators has seen a recent resurgence of interest, with applications spanning quantum optics \cite{intro20,intro21}, nonlinear physics \cite{intro22,intro23,intro25}, molecular chemistry \cite{intro28,intro29}, and quantum chemistry \cite{intro30,intro31}. In modern studies of coupled quantum harmonic oscillators, quantum entanglement \cite{ent1,ent2,ent4,ent5,ent6} and EPR steering \cite{steer2,steer3,steer4,steer5,steer6} attract the greatest interest.\par

Inspired by the work in \cite{steer1}, we introduce and extend a computable method for quantum steering in coupled systems, with a focus on the case of three coupled harmonic oscillators. We employ the geometrical diagonalization approach to reduce the system's degrees of freedom. We aim to understand the steering of three coupled harmonic oscillators and to find the optimal conditions for determining steerable states. The remainder of this paper is organized as follows. In Sec.(\ref{sec2}), we introduce the three-coupled harmonic oscillator system, describe its physical setup, and obtain the corresponding energy spectrum. In Sec.(\ref{sec3}), to obtain the average value, we first identify the Wigner function used to calculate the phase-space fluctuations. In Sec.(\ref{sec4}), we explore quantum steering and focus on its stationary properties within the system, analyzing the conditions that give rise to quantum steering. Finally, in Sec.(\ref{sec5}), we summarize our main findings.

\section{Hamiltonian and Energy spectrum \label{sec2} }
We consider a system of three coupled harmonic oscillators, labeled $x$, $y$, and $z$, where each oscillator is characterized by its own angular frequency $\omega_x$, $\omega_y$, and $\omega_z$, respectively. The Hamiltonian of this system, which describes both the kinetic and potential energy of the oscillators as well as their interactions, is given by the following quadratic Hamiltonian 
\begin{equation}
\mathcal{H} = \frac{1}{2}(\hat{p}_x^{2} +  \hat{p}_y^{2} + \hat{p}_z^{2})  + \frac{1}{2}\omega_x^2  \hat{x}^2 + \frac{1}{2}\omega_y^2  \hat{y}^2 + \frac{1}{2}\omega_z^2  \hat{z}^2 + J_{xy}  \hat{x}  \hat{y} + J_{xz}  \hat{x}  \hat{z} + J_{yz}  \hat{y}  \hat{z} \label{equation1}
\end{equation}
Where the position and momentum operators satisfy $[ \hat{x},  \hat{p}_x] = [ \hat{y},  \hat{p}_y] = [ \hat{z},\hat{p}_z ] = i$, $[ \hat{x}, \hat{y}] =[ \hat{x},  \hat{z}] =[ \hat{z},  \hat{y}] =0$ and $[ \hat{p}_x,  \hat{p}_y] =[ \hat{p}_x,  \hat{p}_z]=[ \hat{p}_y,  \hat{p}_z]=0$. Throughout this paper, we will consider, for simplicity and without loss of generality, $\hbar = m = 1$ \cite{refham1}.
\begin{figure}[H]
    \centering
    \includegraphics[width=0.6\linewidth]{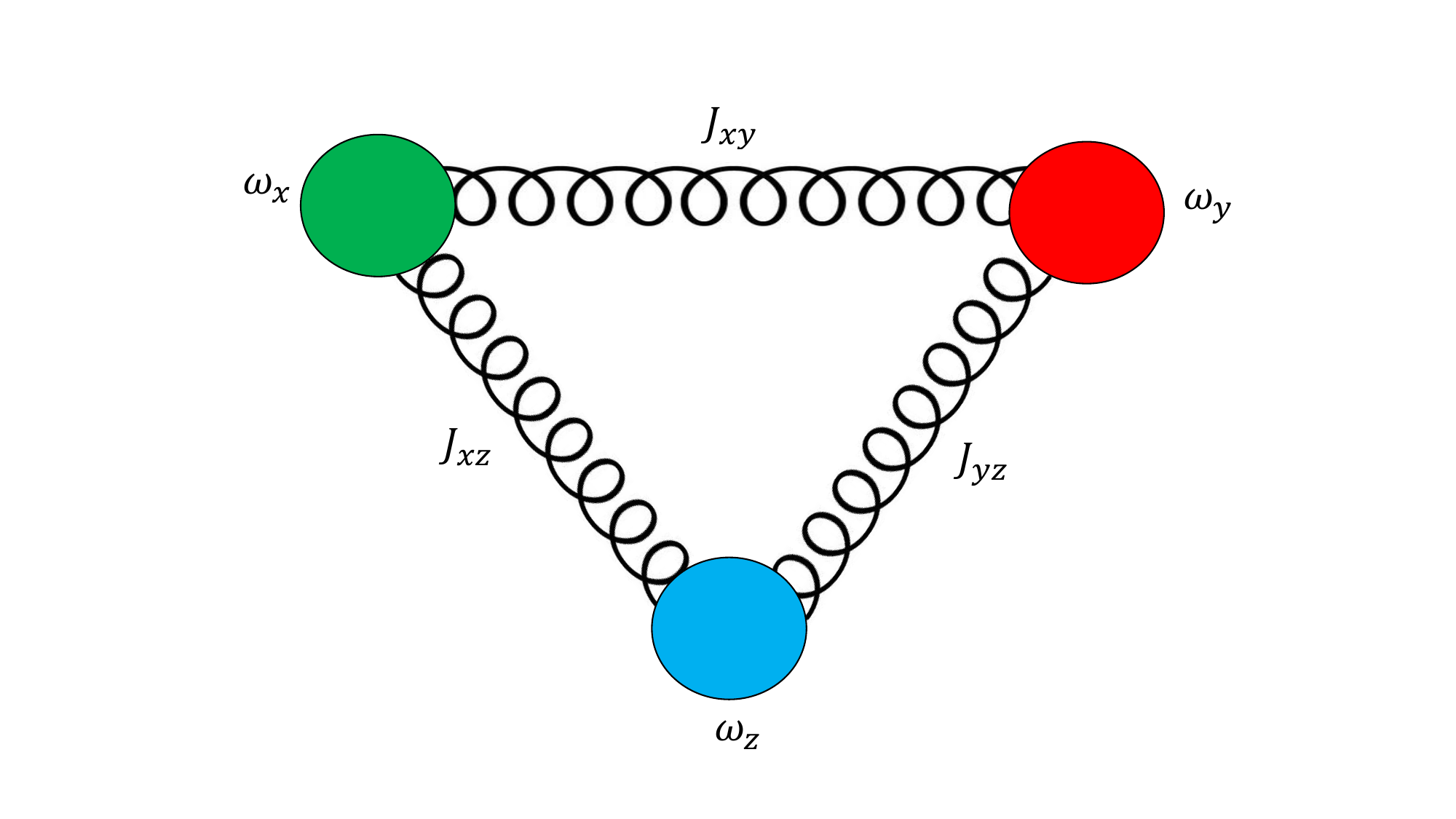}
    \caption{\centering The schematic illustrates three coupled oscillators. Each of the two oscillators, \(\alpha\) and \(\beta\), is coupled via "position-position" interaction \(\alpha \beta\) with a coupling strength of \(J_{\alpha \beta}\), for all \(\alpha \neq \beta\) and \(\alpha, \beta \in \{x,y,z\}\).
    }
    \label{fig:enter-label}
\end{figure}
To diagonalize Eq. (\ref{equation1}), we will harness the Euler unitary transformation defined by \cite{refham2}.
\begin{align}
        \mathbb{R}(\varphi, \Phi, \theta) &= e^{i\varphi \hat{L}_1} e^{i\Phi \hat{L}_2} e^{i\theta \hat{L}_3}\\
        &=\mathbb{R}_1(\varphi)\mathbb{R}_2(\Phi)\mathbb{R}_3(\theta) 
\end{align}
The generators $\hat{L}_k = (\hat{x}_i \hat{p}_j - \hat{x}_j \hat{p}_i)$ fulfill the algebra $[\hat{L}_i, \hat{L}_j] = i\,\epsilon_{ijk}\,\hat{L}_k$ \cite{refham3}, where $\hat{x}_i$ and $\hat{p}_i$ represent the canonical position and momentum operators, $\epsilon_{ijk}$ is the Levi-Civita antisymmetric tensor, and $i$ is the imaginary unit. Therefore, the rotation matrix can be explicitly represented as:
\begin{equation}
    \mathbb{R}(\varphi, \Phi, \theta) = \left( \begin{array}{ccc}
C_\theta C_\Phi & S_\Phi & S_\theta C_\Phi \\
- C_\varphi S_\Phi C_\theta - S_\varphi S_\theta & C_\varphi C_\Phi & - C_\varphi S_\Phi S_\theta + S_\varphi C_\theta \\
S_\varphi S_\Phi C_\theta - C_\varphi S_\theta & - S_\varphi C_\Phi & S_\varphi S_\Phi S_\theta + C_\varphi C_\theta
\end{array} \right)\label{eq5}
\end{equation}
where we have set 
\[
(S_\gamma, C_\gamma) \equiv (\sin \gamma, \cos \gamma), \quad \gamma \in \{ \Phi,\theta, \varphi\}
\]
Then the diagonalized form of the Hamiltonian in Eq. (\ref{equation1}), reduces to 
\begin{equation}
\mathcal{H}_d = \frac{P^2 + Q^2 + F^2}{2} + \frac{1}{2}\vartheta_x^2 X^2 + \frac{1}{2}\vartheta_y^2 Y^2+ \frac{1}{2}\vartheta_z^2 Z^2 \label{eq4}
\end{equation}
where the new position coordinates are given as:
\begin{align}
   X &= x C_\theta C_\Phi + y (-C_\theta C_\varphi S_\Phi - S_\theta S_\varphi) + z (C_\theta S_\varphi S_\Phi - S_\theta C_\varphi)\label{eq1} \\
   Y &= x S_\Phi + y C_\varphi C_\Phi - z S_\varphi C_\Phi \label{eq2}\\
   Z &= x S_\theta C_\Phi + y (C_\theta S_\varphi - S_\theta C_\varphi S_\Phi) + z (S_\theta S_\varphi S_\Phi + C_\theta C_\varphi) \label{eq3}
\end{align}
and their corresponding momenta written as:
\begin{align}
    P &=  p_x C_\theta C_\Phi +p_y (-C_\theta C_\varphi S_\Phi - S_\theta S_\varphi) + p_z (C_\theta S_\varphi S_\Phi - S_\theta C_\varphi) \\
   Q &= p_x S_\Phi + p_y C_\varphi C_\Phi-p_z S_\varphi C_\Phi  \\
   F &=p_x S_\theta C_\Phi + p_y (C_\theta S_\varphi - S_\theta C_\varphi S_\Phi)+ p_z (S_\theta S_\varphi S_\Phi + C_\theta C_\varphi)\label{equation10}
\end{align}
We emphasize that our system involves six degrees of freedom (i.e., three frequencies and three couplings), which significantly increases the complexity of the system. As a result, the study of quantum steering between the three modes becomes a fastidious task. \\
We utilize the geometrical diagonalization approach \cite{2nd_chaptre 2} that constrains Euler angles, which reduces the degrees of freedom of the studied system. In conclusion, we emphasize that the behavior of our three coupled quantum oscillators can be entirely described by a singular mixing angle, $\mu_{\theta}$. It is crucial to emphasize that the relationship between $\mu_{\Phi}$, $\mu_{\varphi}$, and $\mu_{\theta}$ is as follows in this framework:
\begin{equation}
    \mu _{\Phi }=\frac{1-\mu _{\theta }^2-\sqrt{3-\mu _{\theta }^2}}{\sqrt{2+\mu _{\theta }^2-\mu _{\theta }^4}}, \quad \mu _{\varphi }= \frac{1-\mu _{\theta } \sqrt{2-\mu _{\theta }^2}}{\mu _{\theta }^2-1} 
\end{equation}
where $ \mu_{\varphi}=\tan(\varphi)$, $\mu_{\Phi}=\tan(\Phi)$ and $ \mu_{\theta}=\tan(\theta)$\\
We consider the following approximation:
\begin{equation}
 \omega_\alpha\sim \omega_\beta\sim \vartheta_\alpha \sim \vartheta_\beta \sim \vartheta, \qquad \forall \, \alpha\neq \beta,\quad \text{and}\quad \alpha,\beta\in \lbrace x,y,z\rbrace. \label{weak1}
\end{equation}
This approximation assumes that the system possesses a single characteristic frequency, denoted as $\vartheta$. This assumption facilitates our understanding of the quantum steering phenomena in multi-body systems.\\
To go further, we explicit the eigenenergies  
\begin{equation}
E_{(n,m,l)}= \vartheta_x \left( n + \frac{1}{2} \right) + \vartheta_y \left( m + \frac{1}{2} \right)+\vartheta_z \left( l + \frac{1}{2} \right) \simeq \vartheta \left( n+m+l + \frac{3}{2} \right)
\end{equation}
and the eigenfunctions as
\begin{equation}
\hspace{-3cm}
\begin{aligned}
\Psi_{(n,m,l)}(X, Y,Z) = \frac{1}{\sqrt{2^{n+m+l} n! m! l!}} \left( \frac{\vartheta}{\pi} \right)^{\frac{3}{4}} 
\exp\left({-\frac{\vartheta}{2} (X^2+Y^2+Z^2)}\right) 
H_n(\sqrt{\vartheta} X) H_m(\sqrt{\vartheta} Y)H_l(\sqrt{\vartheta} Z),
\end{aligned}
\end{equation}
where $n,m$ and $l$ stand for quantum numbers, and the special functions $H_{n,m,l}$ are the Hermite orthogonal polynomials.
\section{Wigner function and The average values \label{sec3} }
\subsection{Wigner function}
Among many representations of the quantum state,
The Wigner function offers an appealing possibility to describe quantum phenomena using the classical-like concept of phase space \cite{ref7,ref8}. The Wigner function provides
complete information on the state of a system \cite{ref8}. The Wigner function corresponding to  our diagonalized Hamiltonian (\ref{eq4}) is separable, such that
\begin{equation}
W_{(n,m,l)}(X, P; Y, Q;Z,F) = W_n(X, P) \times W_m(Y, Q)\times W_l(Z, F),\label{eq21}
\end{equation}
where the marginal Wigner functions are
\begin{align}
W_n(X, P) &= \frac{1}{\pi} \int \Psi_n^*(X + \mathbb {X}) \Psi_n(X - \mathbb {X}) \exp\left({2iP\mathbb {X}}\right) d\mathbb {X} \notag\\
          &= \frac{(-1)^n}{\pi} \exp\left({-\frac{1}{\vartheta_x} (\vartheta_x^2 X^2 + P^2)}\right)\mathcal{L}_n \left[ \frac{2}{\vartheta_x} (\vartheta_x^2 X^2 + P^2) \right],\label{eq22}\\
W_m(Y, Q) &= \frac{1}{\pi} \int \Psi_m^*(Y + \mathbb {Y}) \Psi_m(Y -  \mathbb {Y}) \exp\left({2iQ \mathbb {Y}}\right) d \mathbb {Y} \notag \\
          &= \frac{(-1)^m}{\pi} \exp\left({-\frac{1}{\vartheta_y} (\vartheta_y^2 Y^2 + Q^2)}\right) \mathcal{L}_m \left[ \frac{2}{\vartheta_y} (\vartheta_y^2 Y^2 + Q^2) \right],\label{eq23}\\
W_l(Z, F) &= \frac{1}{\pi} \int \Psi_l^*(Z + \mathbb {Z}) \Psi_l(Z - \mathbb {Z}) \exp\left({2iF\mathbb {Z}}\right) d\mathbb {Z} \notag\\
          &= \frac{(-1)^l}{\pi} \exp\left({-\frac{1}{\vartheta_z} (\vartheta_z^2 Z^2 + F^2)}\right) \mathcal{L}_l \left[ \frac{2}{\vartheta_z} (\vartheta_z^2 Z^2 + F^2) \right],\label{eq24}
\end{align}
where $\mathcal{L}_n(x)$ is Laguerre polynomials \cite{ref9}. To go further,  we harness Rodrigues' formula for Laguerre polynomials \cite{ref9, ref10}, 
\begin{equation}
\mathcal{L}_n(x) = \frac{1}{n!} \frac{d^n}{du^n} \left( \frac{ \exp\left({-\frac{x u}{1-u}}\right)}{1-u} \right) \bigg|_{u=0}. \label{eq25}
\end{equation}
Under the realistic assumption in Eq. (\ref{weak1}), we end up with the Wigner function associated with our Hamiltonian in Eq. (\ref{eq4})
\begin{align}
\hspace{-1cm} W_{(n,m,l)}(X, P; Y, Q; Z, F) &=\mathbb{R}_{n,l,m} \left( -\frac{\exp \left(\frac{1}{\vartheta} \left[\frac{(u+1)(P^2+ X^2 \vartheta^2)}{(u-1)} + \frac{(s+1)(Q^2 +Y^2 \vartheta^2)}{(s-1) }+\frac{(v+1) \left(F^2 + Z^2 \vartheta^2 \right)}{(v-1)}  \right]\right)}{(s-1) (u-1) (v-1)} \right)\Bigg|_{u,s,v=0}, \label{equation23}
\end{align}
 where the operator 
 \begin{equation}
 \mathbb{R}_{n,l,m}=\frac{ (-1)^{l+m+n}}{\pi^3 l! m! n!} \frac{d^n}{du^n} \frac{d^m}{ds^m} \frac{d^l}{dv^l}.
 \end{equation}
 The Wigner function will be crucial for the upcoming analysis of quantum steering in our system. Specifically, quantum steerability depends on the average values of the creation and annihilation operators associated with each oscillator.
\subsection{The average values}
To analyze the quantum fluctuations for our system, the expectation value of an operator $\mathcal{A}$ is given by the following formula \cite{ref11}.
\begin{equation}
\langle \mathcal{A} \rangle=\int_{\mathbb{R}^6} \mathcal{A}\times {W}_{n,m}(x,p_x;y,p_y;z,p_z) dx dp_x dy dp_y dz dp_z
\end{equation}
We define the average values of positions and momenta in \textbf{(First Appendix I)}. By using the expectation values, it is straightforward to show the following sum rules
\begin{align}
   \langle x^2 \rangle +\langle y^2 \rangle+\langle z^2 \rangle &= \frac{1}{2} \left(\frac{2 l+1}{\vartheta _z}+\frac{2 m+1}{\vartheta _y}+\frac{2 n+1}{\vartheta _x}\right)= \frac{1}{\vartheta} \left( n+ m+ l+\frac{3}{2}\right)\\
    \langle p_x^2 \rangle +\langle p_y^2 \rangle+\langle p_z^2 \rangle &=\frac{1}{2} \Big((2 l+1) \vartheta _z+(2 m+1) \vartheta _y+(2 n+1) \vartheta _x\Big)=\vartheta \left( n+ m+ l+\frac{3}{2}\right)
\end{align}
We would like to highlight that these results are consistent with those previously presented in \cite{refrence12}. Moreover, we can show that the expectation values obey the following relation:
\begin{align}
   Cov(x^2;p_y^2)=&Cov(y^2;p_x^2),\qquad 
Cov(z^2;p_x^2)=Cov(x^2;p_z^2), \qquad Cov(y^2;p_z^2)=Cov(z^2;p_y^2)
\end{align}
where, the covariance between two observables is $Cov(X,Y)=\langle XY\rangle-\langle X\rangle \langle Y\rangle$.
\section{Quantum Steering and  Numerical Results \label{sec4} }
\subsection{Quantification of quantum steering}
Quantum steering refers to the fact that, in a bipartite scenario, one of the parties can change the state
of the other distant party by applying local measurements \cite{intro4}. In other words, steering is a quantum mechanical process that enables one party, \textbf{A}, to change (to "\textit{steer}") the state of another distant party, \textbf{B} \cite{ref12,ref13} in a way that cannot be explained by classical correlations \cite{steer1}.
The creation and annihilation operators are 
\begin{equation}
 (a_\beta^{\dagger})^{\dagger} = a_{\beta} = \sqrt{\frac{\omega_\beta}{2}} \beta + \frac{i}{\sqrt{2\,\omega_\beta}}p_{\beta} \qquad \beta \in \{x,y,z\}
\end{equation}
The parameter that allows for the detection of quantum steering and quantifying the steerability of all the possible cases \cite{reference16}, can be expressed as 
 \begin{eqnarray}
      S_{{\alpha} \rightarrow \beta}^{(n, m,l)} &=&max \left[ \left|\langle a_{\alpha} a_\beta^{\dagger}\rangle \right|^2  - \left< a_\beta^{\dagger} a_\beta \left( a_{\alpha}^{\dagger} a_{\alpha} + \frac{1}{2} \right) \right>, 0\right] \\ 
     S_{\beta \rightarrow {\alpha}}^{(n, m,l)} &=&max \left[ \left|\langle a_{\alpha} a_\beta^{\dagger}\rangle \right|^2 - \left< a_{\alpha}^{\dagger} a_{\alpha} \left( a_\beta^{\dagger} a_\beta + \frac{1}{2} \right) \right>, 0\right]
 \end{eqnarray}
Expanding the steering between $x$ and $y$ gives :
\begin{align}
    S_{x \rightarrow y}^{(n, m,l)} = max \Bigg[&\frac{w_x w_y}{4}\langle xy \rangle^2 + \frac{1}{4 w_x w_y}\langle p_x p_y \rangle^2 + \frac{1}{2} \langle x y \rangle \langle  p_x p_y \rangle - \frac{w_x w_y}{4} \langle x^2y^2 \rangle \notag\\
    &\phantom{=}-\frac{w_x}{4 w_y} \langle x^2 p_y^2 \rangle - \frac{w_y}{4w_x} \langle p_x^2y^2 \rangle- \frac{1}{4w_x w_y} \langle p_x^2 p_y^2 \rangle + \frac{w_x}{4} \langle x^2 \rangle+ \frac{1}{4w_x} \langle p_x^2 \rangle ,0\Bigg] 
\\
    S_{y \rightarrow x}^{(n, m,l)} = max\Bigg[&\frac{w_x w_y}{4}\langle xy \rangle^2 + \frac{1}{4 w_x w_y}\langle p_x p_y \rangle^2 + \frac{1}{2} \langle xy \rangle \langle  p_x p_y \rangle - \frac{w_x w_y}{4} \langle x^2y^2 \rangle\notag \\
    &\phantom{=}-\frac{w_x}{4 w_y} \langle x^2 p_y^2 \rangle - \frac{w_y}{4w_1} \langle p_x^2y^2 \rangle- \frac{1}{4w_1 w_y} \langle p_x^2 p_y^2 \rangle + \frac{w_y}{4} \langle y^2 \rangle+ \frac{1}{4w_2} \langle p_y^2 \rangle,0 \Bigg]
\end{align}
\subsection{ Numerical Results and Discussions}
In this section, we will analyze the structure of quantum steering within three particles. Quantum steering in all directions $(\alpha \leftrightarrow \beta)$ is defined in \textbf{(Second Appendix II)}. This expression's form shows that the quantum steering between the three oscillators is dependent on the quantum states as well as the mixing angles. Additionally, we recall that steering is not symmetric. Then, its asymmetry is defined as :
\begin{align}
    \Delta S_{xy}=& \lvert S_{x \rightarrow y} - S_{y \rightarrow x} \rvert \notag\\
    =&\frac{(m-n) \left(\mu _{\theta }^2 \mu _{\varphi }^2 \mu _{\Phi }^2+\mu _{\Phi }^2-1\right)}{2 \left(\mu _{\theta }^2+1\right) \left(\mu _{\varphi }^2+1\right) \left(\mu _{\Phi }^2+1\right)}+\frac{(l-n) \left(2 \mu _{\theta } \mu _{\varphi } \sqrt{\mu _{\Phi }^2+1} \mu _{\Phi }+\mu _{\theta }^2 \mu _{\varphi }^2\right)}{2 \left(\mu _{\theta }^2+1\right) \left(\mu _{\varphi }^2+1\right) \left(\mu _{\Phi }^2+1\right)}-\frac{(l-m)\mu _{\theta }^2  \left(\mu _{\Phi }^2-1\right)}{2 \left(\mu _{\theta }^2+1\right) \left(\mu _{\varphi }^2+1\right) \left(\mu _{\Phi }^2+1\right)}\\
    &\phantom{=}+\frac{\mu _{\varphi }^2 \left((m-l) \mu _{\Phi }^2-l+n\right)}{2 \left(\mu _{\theta }^2+1\right) \left(\mu _{\varphi }^2+1\right) \left(\mu _{\Phi }^2+1\right)}\notag
\end{align}
We discover the asymmetry in the system, indicating that the subsystems cannot control each other equally. This shows that one subsystem may have a greater impact or control over another.\\
We also identify an equation that connects the trade-offs among all subsystems as
\begin{equation}
    \Delta S_{xy}-\Delta S_{xz}-\Delta S_{zy}=0
\end{equation}
We present here a clear visual representation of quantum steering for all directions and for various quantum numbers $(n,m,l)$ ranging from $0$ to $100$ and $\mu_{\theta} \in [-1,1]$. The plotted results show that the steering depends on the direction of quantum steering and the mixing angle $\mu_{\theta}$. This allows us to observe the pattern and the behavior of the quantum steering in all possible directions.\par 

\begin{figure}[H]
    \centering
    \includegraphics[width=0.32\linewidth]{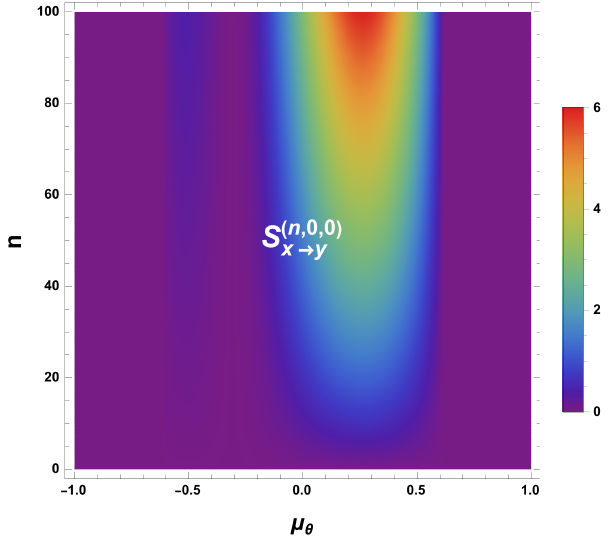}
    \includegraphics[width=0.32\linewidth]{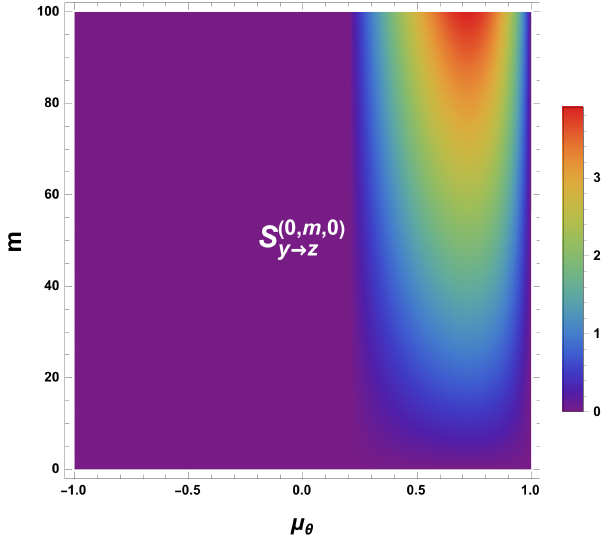}
    \includegraphics[width=0.32\linewidth]{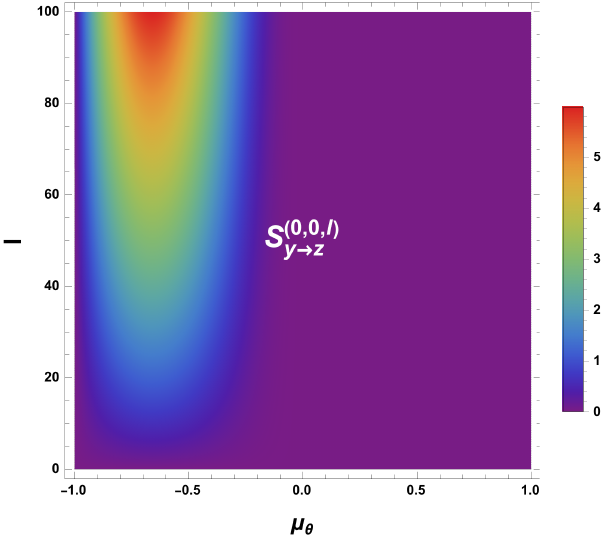}
    \caption{\centering The evolution of quantum steering $S^{(n,m,l)}_{x \rightarrow y}$ and $S^{(n,m,l)}_{y \rightarrow x}$ versus quantum numbers $(n,m,l)$, and $\mu_{\theta}$ \label{fig2} }
\end{figure}

In {\sf Figure} \ref{fig2}, the steering \((x\leftrightarrow y)\) exhibit a highest values under these considerations reaches $\simeq 6$ at a specific values of $\mu_{\theta}$. As a general observation, we note that quantum steering is an increasing function of the quantum numbers for a given value of the mixing angle $\mu_\theta$ in both directions, indicating that increased excitation levels of the system lead to stronger steering effects. Here we present all the states where steering occurs. Additionally, the steering of the other states in subsystem \((x\leftrightarrow y)\) are equal to zero; 
\(( S^{(0,m,0)}_{x \rightarrow y}(\theta)=S^{(0,0,l)}_{x \rightarrow y}(\theta)=S^{(n,0,0)}_{y \rightarrow x}(\theta)=0)\).

The plots reveal clear regions of strong and weak quantum correlation as a function of $(n,m,l)$ and the mixing angle $\mu_{\theta}$. We observe that the quantum steering appears for specific values of $\mu_\theta$, and the steering $S^{(n,0,0)}_{x \rightarrow y}$ and $S^{(0,m,0)}_{y \rightarrow x}$ occur for positive values of $\mu_\theta$. On the other hand, the steering $S^{(0,0,l)}_{y \rightarrow x}$ arises only for negative $\mu_\theta$. Additionally, higher quantum numbers strengthen the quantum steering in both directions, while the ground state $(0,0,0)$ shows no correlation. Hence, the quantum steering is given as:  
\begin{align}
    S^{(n,0,0)}_{x\to y}&= -\frac{n \left(\mu _{\theta }^2 \left(\mu _{\Phi }^2+1\right)+\mu _{\Phi }^2-1\right)}{2 \left(\mu _{\theta }^2+1\right)^2 \left(\mu _{\varphi }^2+1\right) \left(\mu _{\Phi }^2+1\right)^2}\bigg[ \mu _{\theta }^2 \mu _{\varphi }^2 \left(\mu _{\Phi }^2+1\right)+2 \mu _{\theta } \mu _{\varphi }\mu _{\Phi } \sqrt{\mu _{\Phi }^2+1} +\mu _{\Phi }^2\bigg]\\
   S^{(0,m,0)}_{y\to x}&= -\frac{m \mu _{\Phi }^2 \left(\mu _{\varphi }^2 \left(\mu _{\Phi }^2+1\right)+\mu _{\Phi }^2-1\right)}{2 \left(\mu _{\varphi }^2+1\right) \left(\mu _{\Phi }^2+1\right)^2}\\
   S^{(0,0,l)}_{y\to x}&=-\frac{l \mu _{\theta }^4 \left(\mu _{\Phi }^2\left(\mu _{\varphi }^2-1\right)+\mu _{\varphi }^2+1\right)}{2 \left(\mu _{\theta }^2+1\right)^2\left(\mu _{\varphi }^2+1\right) \left(\mu _{\Phi }^2+1\right)^2}-\frac{l \mu _{\theta }^2 \left(4 \mu _{\theta } \mu _{\varphi } \mu _{\Phi } \sqrt{\mu _{\Phi }^2+1}-\left(\mu _{\varphi }^2-1\right) \left(\mu _{\Phi }^2+1\right)\right)}{2 \left(\mu _{\theta }^2+1\right)^2\left(\mu _{\varphi }^2+1\right) \left(\mu _{\Phi }^2+1\right)^2}
\end{align}
\begin{figure}[H]
    \centering
    \includegraphics[width=0.32\linewidth]{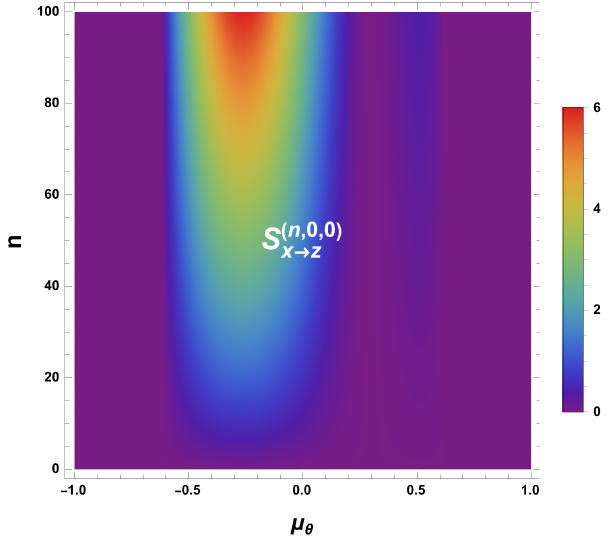}
    \includegraphics[width=0.32\linewidth]{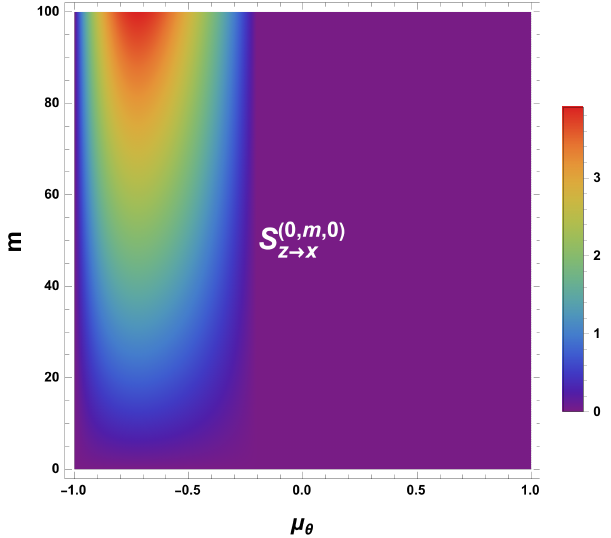}
    \includegraphics[width=0.32\linewidth]{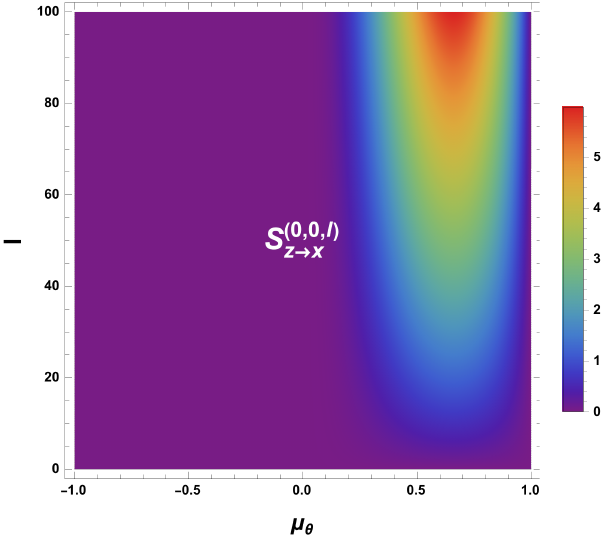}
    \caption{\centering  The evolution of quantum steering $S^{(n,m,l)}_{x \rightarrow z}$ and $S^{(n,m,l)}_{z \rightarrow x}$ versus quantum numbers $(n,m,l)$, and $\mu_{\theta}$ \label{fig3}}
\end{figure}
{\sf Figure} \ref{fig3} displays the amount of quantum steering in both directions $(x \rightarrow z)$ and $(z \rightarrow x)$ versus $\theta$ for the states $(n,0,0)$, $(0,m,0)$, and $(0,0,l)$. This graph indicates that the steering increases as the quantum numbers become more highly excited. And the steering appears for certain values of $\mu_\theta$ in each state; the parameter $\mu_\theta$ effectively tunes the strength of this steering, from complete separability to maximal correlation. We observe that the steering of subsystem $(x \leftrightarrow z)$ exhibits a symmetric behavior like the one in $(x \leftrightarrow y)$. We can express this symmetrical behavior as 
\begin{equation}
    S^{(n,m,l)}_{x\to z}(\theta)=S^{(n,m,l)}_{x\to y}(-\theta) \quad\text{and}\quad S^{(n,m,l)}_{z\to x}(\theta)=S^{(n,m,l)}_{y\to x}(-\theta)
\end{equation}
The steering of the other states in the subsystem $(x\leftrightarrow z)$ is equal to zero. This can be expressed as $( S^{(0,m,0)}_{x \rightarrow z}=S^{(0,0,l)}_{x \rightarrow z}=S^{(n,0,0)}_{z \rightarrow x}=0)$.
\begin{figure}[H]
      \centering
    \includegraphics[width=0.32\linewidth]{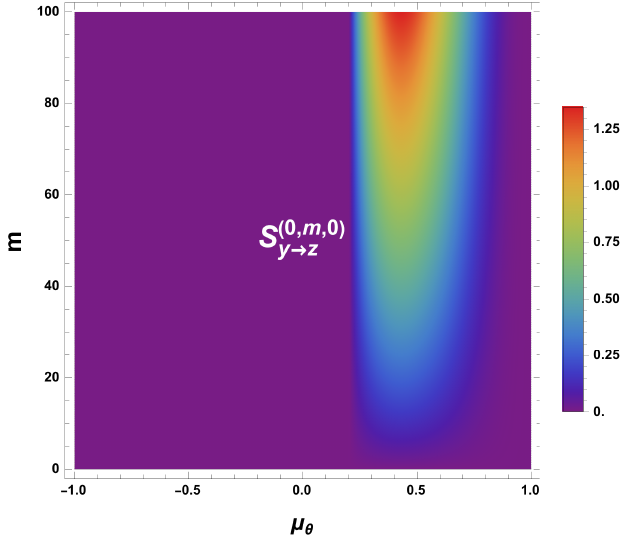}
    \includegraphics[width=0.32\linewidth]{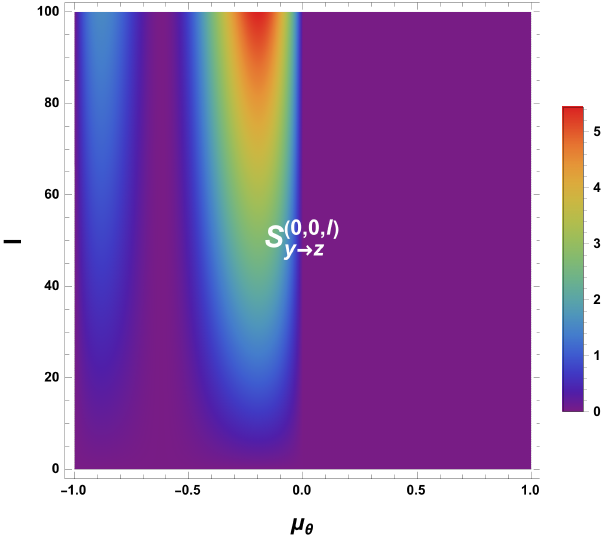}
     \par\medskip
    \includegraphics[width=0.32\linewidth]{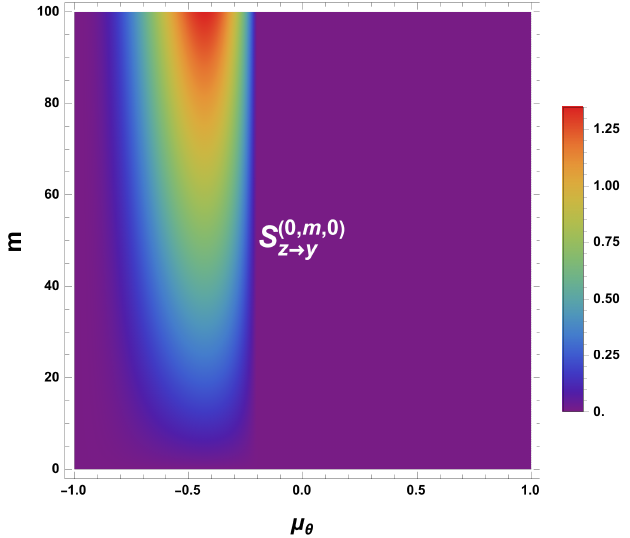}
    \includegraphics[width=0.32\linewidth]{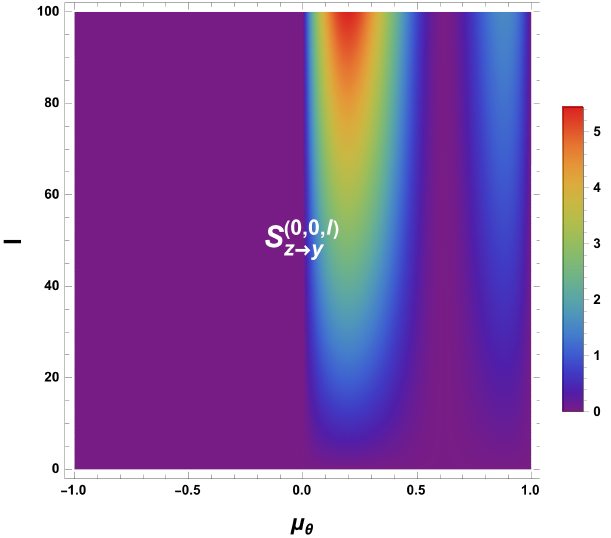}
     \caption{\centering  The evolution of quantum steering $S^{(n,m,l)}_{y \rightarrow z}$ and $S^{(n,m,l)}_{z \rightarrow y}$ versus quantum numbers $(n,m,l)$, and $\mu_{\theta}$}
    \label{fig:quantum-steering}\label{fig4}
\end{figure}
In {\sf Figure} \ref{fig4}, we illustrate the quantum steering in the directions ($y \rightarrow z$) and ($z \rightarrow y$) versus quantum numbers $m$ and $l$. And the mixing angle $\mu_{\theta}$. The steering appears for the positive values of $\mu_\theta$ in both directions ($y \rightarrow z$) and ($z \rightarrow y$) for the states $(0,m,0)$ and $(0,0,l)$, respectively. On the other hand, the steering occurs for negative values of $\mu_\theta$ in the directions ($y \rightarrow z$) and ($z \rightarrow y$) for the states $(0,0,l)$ and $(0,m,0)$, respectively.  Additionally, the highest values of steering between $y$ and $z$ are observed for the states $(0,0,l)$, reaching a maximum of approximately $\simeq 5.4$. We notice that the quantum steering in subsystem $(y \leftrightarrow z)$ shows an intrinsic symmetry between $(y \rightarrow z)$ and $(z \rightarrow y)$ for the same states, which we can express as
\begin{equation}
        S^{(n,m,l)}_{y\to z}(\theta)=S^{(n,m,l)}_{z\to y}(-\theta) 
\end{equation}
As a result, we give the general expression of the quantum steering in the direction $(y \rightarrow z)$ as
\begin{align}
    S^{(0,m,0)}_{y\to z}&=-\frac{m \mu _{\varphi }^2 \left(\mu _{\varphi }^2 \left(\mu _{\Phi }^2+1\right)+\mu _{\Phi }^2-1\right)}{2 \left(\mu _{\varphi }^2+1\right)^2 \left(\mu _{\Phi }^2+1\right)^2}\\
    S^{(0,0,l)}_{y\to z}&=-\frac{l \mu _{\theta }^4 \mu _{\varphi }^2 \mu _{\Phi }^2 \left(\left(\mu _{\varphi }^2-1\right) \mu _{\Phi }^2+\mu _{\varphi }^2+1\right)}{2 \left(\mu _{\theta }^2+1\right)^2 \left(\mu _{\varphi }^2+1\right)^2 \left(\mu _{\Phi }^2+1\right)^2}-\frac{l \mu _{\theta }^3 \mu _{\varphi } \mu _{\Phi } \left(\left(3 \mu _{\varphi }^2-1\right) \mu _{\Phi }^2+\mu _{\varphi }^2+1\right)}{\left(\mu _{\theta }^2+1\right)^2 \left(\mu _{\varphi }^2+1\right)^2 \left(\mu _{\Phi }^2+1\right){}^{3/2}}\notag\\
    &\phantom{=}+\frac{l \mu _{\theta }^2 \left(\mu _{\varphi }^2 \left(\left(\mu _{\varphi }^2-10\right) \mu _{\Phi }^2-1\right)+\mu _{\Phi }^2-1\right)}{2 \left(\mu _{\theta }^2+1\right)^2 \left(\mu _{\varphi }^2+1\right)^2 \left(\mu _{\Phi }^2+1\right)}+\frac{l \mu _{\theta } \mu _{\varphi } \left(\mu _{\varphi }^2-3\right) \mu _{\Phi }}{\left(\mu _{\theta }^2+1\right)^2 \left(\mu _{\varphi }^2+1\right)^2 \sqrt{\mu _{\Phi }^2+1}}+\frac{l \left(\mu _{\varphi }^2-1\right)}{2 \left(\mu _{\theta }^2+1\right)^2 \left(\mu _{\varphi }^2+1\right)^2}
\end{align}
The table below displays the quantum steering that is set to zero for a single excitation level in all directions.
\begin{table}[H]
    \centering
    \label{tab:steering_excitation}
    \begin{tabular}{|c|c|c|c|}
        \hline
        \textbf{Steering Direction} & \textbf{$(n,0,0)$} & \textbf{$(0,m,0)$} & \textbf{$(0,0,l)$} \\
        \hline
        $x \rightarrow y$ & $\neq 0$ & $0$ & $0$ \\
        $y \rightarrow x$ & $0$ & $\neq 0$ & $\neq 0$ \\
        \hline
        $x \rightarrow z$ & $\neq 0$ & $0$ & $0$ \\
        $z \rightarrow x$ & $0$ & $\neq 0$ & $\neq 0$ \\
        \hline
        $y \rightarrow z$ & $0$ & $\neq 0$ & $\neq 0$ \\
        $z \rightarrow y$ & $0$ & $\neq 0$ & $\neq 0$ \\
        \hline
    \end{tabular}
    \caption{Steering conditions for different excitation directions.}
\end{table}
The quantum steering in both directions, $(x \rightarrow y)$ and $(x \rightarrow z)$, occurs only in n-level excitation. This is in contrast to $(y \rightarrow x)$,$(z \rightarrow x)$, $(y \rightarrow z)$, and $(z \rightarrow y)$, where the steering vanishes in n-level excitation. These results exhibit the importance of the special conditions on \( l \), \( n \), and \( m \) in deciding if steering is possible between the corresponding directions. It is important to point out that all calculated quantum steering values are completely asymmetric and not equivalent, i.e., $S^{(n,m,l)}_{\alpha \rightarrow \beta} \neq S^{(n,m,l)}_{\beta \rightarrow \alpha}$ with $\alpha,\beta \in \{x, y, z\}$, for different quantum numbers $n$, $m$ and $l$.\\
\begin{figure}[H]
    \centering
    \includegraphics[width=0.32\linewidth]{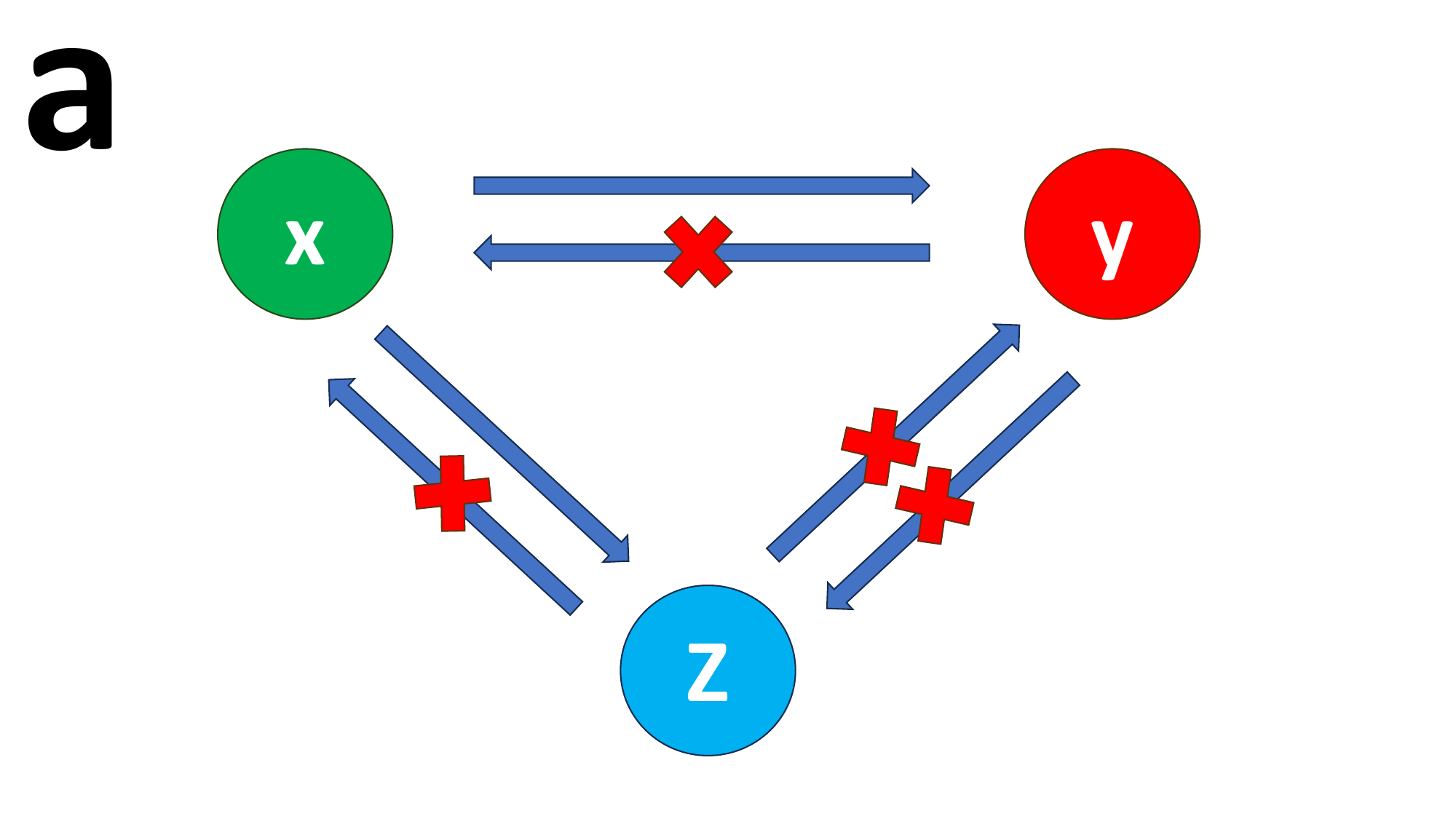}
    \includegraphics[width=0.32\linewidth]{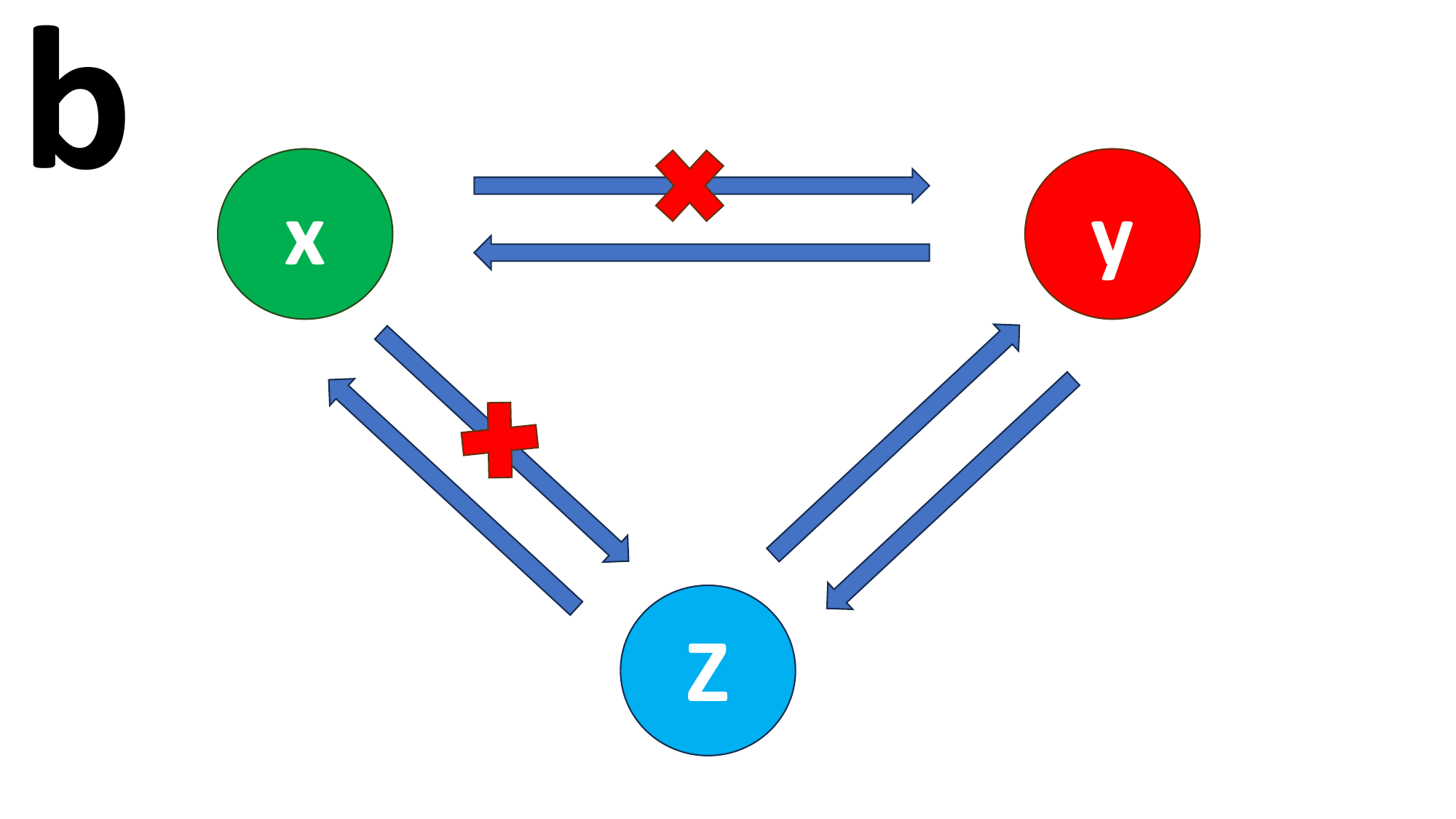}
    \caption{\centering Configurations of the quantum steering correlations among three oscillators. \textbf{a)}. for the states $(n,0.0)$. \textbf{b)}. for the states $(0,m,0)$ and $(0,0,l)$.}\label{fig5}
    \label{fig:enter-label}
\end{figure}
The {\sf Figure} \ref{fig5} exhibits the quantum steering among three quantum oscillators for different states $(n,0,0)$, $(0,m,0)$, and $(0,0,l)$. Panel \textbf{(a)} illustrates the steering for the states $(n,0,0)$, where the steering $(y \leftrightarrow z)$ vanishes. The oscillator $x$ can simultaneously steer both of the other oscillators, $y$ and $z$, at the same time. While panel \textbf{(b)} shows that the states $(0,m,0)$ and $(0,0,l)$ produce identical steering configurations, indicating an intrinsic symmetry between oscillators $y$ and $z$. This figure demonstrates how the directionality and topology of steerable correlations are determined by the excitation locale rather than the excitation magnitude.\\

\section{Conclusion \label{sec5} }
We conducted an in-depth examination of quantum steering between three coupled harmonic oscillators using the \emph{geometrical diagonalization} approach for minimizing the degree of freedom related to our system. We derived analytical expressions for quantum steering of all directions, using the Wigner function in phase space rather than the Schmidt decomposition. Our research shows how quantum steering in a system of three coupled quantum oscillators is highly sensitive to the mixing angle and excitation level. Conversely, the ground state (0,0,0) exhibits no steering at all in any direction. In addition, the directionality and topology of steerable correlations are regulated by the location of the excitation rather than its magnitude. We also observe symmetric steering behavior under equivalent excitation conditions between oscillators $x$, $y$, and $z$. Mathematically, it can be represented by $S^{(n,m,l)}_{x\to z}(\theta)=S^{(n,m,l)}_{x\to y}(-\theta)$,  $S^{(n,m,l)}_{z\to x}(\theta)=S^{(n,m,l)}_{y\to x}(-\theta)$, and $S^{(n,m,l)}_{y\to z}(\theta)=S^{(n,m,l)}_{z\to y}(-\theta)$. In addition, we demonstrate that symmetric steering cannot occur between the same two particles in the three-coupled harmonic oscillator system. Therefore, we write $S^{(n,m,l)}_{\alpha \rightarrow \beta} \neq S^{(n,m,l)}_{\beta \rightarrow \alpha}$ with $\alpha,\beta \in \{x, y, z\}$.

Furthermore, we see that a single particle can simultaneously steer the other two particles in such a system. These results show that adjusting the mixing angle and excitation level is crucial for strong quantum steering. Additionally, they provide a more profound understanding of the behavior of quantum steering in three coupled harmonic oscillators. Understanding quantum steering may enable enhanced quantum control, optimization of information processing within the system, and advancements in quantum technologies. 

\section*{First Appendix I: AVERAGE VALUES OF POSITIONS AND MOMENTA }
We consider in detail the average values of positions and momenta as follows. 
\allowdisplaybreaks % Enables equations to break across pages
\begin{align}
 \langle x^2 \rangle &=\frac{1}{2} \left[\frac{(2 l+1) S^2_{\theta } C^2_{\Phi}}{\vartheta _z}+\frac{(2 m+1) S^2_{\Phi}}{\vartheta _y}+\frac{(2 n+1) C^2_{\theta } C^2_{\Phi}}{\vartheta _x}\right],  \\
 \langle y^2 \rangle &=\frac{1}{2} \bigg[\frac{(2 l+1) (C_{\theta } S_{\varphi}-S_{\theta } C_{\varphi} S_{\Phi})^2}{\vartheta _z}+\frac{(2 m+1) C^2_{\varphi} C^2_{\Phi}}{\vartheta _y}+\frac{(2 n+1) (C_{\theta } C_{\varphi} S_{\Phi}+S_{\theta } S_{\varphi})^2}{\vartheta _x}\bigg],\\
 \langle z^2 \rangle &=\frac{1}{2} \bigg[\frac{(2 l+1) (S_{\theta } S_{\varphi} S_{\Phi}+C_{\theta } C_{\varphi})^2}{\vartheta _z}+\frac{(2 m+1) S^2_{\varphi} C^2_{\Phi}}{\vartheta _y}+\frac{(2 n+1) (S_{\theta } C_{\varphi}-C_{\theta } S_{\varphi} S_{\Phi})^2}{\vartheta _x}\bigg],   \\
 \langle p_x^2 \rangle  &=\frac{1}{2} \bigg[C^2_{\Phi} \left((2 l+1) S^2_{\theta } \vartheta _z+(2 n+1) C^2_{\theta } \vartheta _x\right)+(2 m+1) S^2_{\Phi} \vartheta _y\bigg],  \\
 \langle p_y^2 \rangle  &=\frac{1}{2} \bigg[(2 l+1) \vartheta _z (C_{\theta } S_{\varphi}-S_{\theta } C_{\varphi} S_{\Phi})^2+(2 m+1) C^2_{\varphi} C^2_{\Phi} \vartheta _y+(2 n+1) \vartheta _x (C_{\theta } C_{\varphi} S_{\Phi}+S_{\theta } S_{\varphi})^2\bigg],  \\
 \langle p_z^2 \rangle  &=\frac{1}{2} \bigg[(2 l+1) \vartheta _z (S_{\theta } S_{\varphi} S_{\Phi}+C_{\theta } C_{\varphi})^2+(2 m+1) S^2_{\varphi} C^2_{\Phi} \vartheta _y+(2 n+1) \vartheta _x (S_{\theta } C_{\varphi}-C_{\theta } S_{\varphi} S_{\Phi})^2\bigg],  \\
 \langle x y \rangle &=\frac{1}{2} C_{\Phi} \bigg[\frac{(2 l+1) S_{\theta } (C_{\theta } S_{\varphi}-S_{\theta } C_{\varphi} S_{\Phi})}{\vartheta _z}+\frac{(2 m+1) C_{\varphi} S_{\Phi}}{\vartheta _y}-\frac{(2 n+1) C_{\theta } (C_{\theta } C_{\varphi} S_{\Phi}+S_{\theta } S_{\varphi})}{\vartheta _x}\bigg],  \\
 \langle x z \rangle &=\frac{1}{2} C_{\Phi} \bigg[\frac{(2 l+1) S_{\theta} (S_{\theta} S_{\varphi} S_{\Phi}+C_{\theta} C_{\varphi})}{\vartheta _z}-\frac{(2 m+1) S_{\varphi} S_{\Phi}}{\vartheta _y}+\frac{(2 n+1) C_{\theta} (C_{\theta} S_{\varphi} S_{\Phi}-S_{\theta} C_{\varphi})}{\vartheta _x}\bigg],  \\
 \langle y z \rangle &= \frac{1}{2} \bigg[\frac{(2 l+1) (S_{\theta} S_{\varphi} S_{\Phi}+C_{\theta} C_{\varphi}) (C_{\theta} S_{\varphi}-S_{\theta} C_{\varphi} S_{\Phi})}{\vartheta _z}-\frac{(2 m+1) S_{\varphi} C_{\varphi} C^2_{\Phi}}{\vartheta _y}\notag\\
 &\phantom{=}+\frac{(2 n+1) (C_{\theta} C_{\varphi} S_{\Phi}+S_{\theta} S_{\varphi}) (S_{\theta} C_{\varphi}-C_{\theta} S_{\varphi} S_{\Phi})}{\vartheta _x}\bigg],\\
 \langle p_x p_y \rangle &=\frac{1}{2} C_{\Phi} \bigg[(2 l+1) S_{\theta} \vartheta _z (C_{\theta} S_{\varphi}-S_{\theta} C_{\varphi} S_{\Phi})+(2 m+1) C_{\varphi} S_{\Phi} \vartheta _y-\left((2 n+1) C_{\theta} \vartheta _x (C_{\theta} C_{\varphi} S_{\Phi}+S_{\theta} S_{\varphi})\right)\bigg] \\
 \langle p_x p_z \rangle &=\frac{1}{2} C_{\Phi} \bigg[(2 l+1) S_{\theta} \vartheta _z (S_{\theta} S_{\varphi} S_{\Phi}+C_{\theta} C_{\varphi})-(2 m+1) S_{\varphi} S_{\Phi} \vartheta _y+(2 n+1) C_{\theta} \vartheta _x (C_{\theta} S_{\varphi} S_{\Phi}-S_{\theta} C_{\varphi})\bigg], \\
 \langle p_y p_z \rangle &=\frac{1}{2} \bigg[(2 l+1) \vartheta _z (S_{\theta} S_{\varphi} S_{\Phi}+C_{\theta} C_{\varphi}) (C_{\theta} S_{\varphi}-S_{\theta} C_{\varphi} S_{\Phi})-\left((2 m+1) S_{\varphi} C_{\varphi} C^2_{\Phi} \vartheta _y\right)\notag\\
 &\phantom{=}+(2 n+1) \vartheta _x (C_{\theta} C_{\varphi} S_{\Phi}+S_{\theta} S_{\varphi}) (S_{\theta} C_{\varphi}-C_{\theta} S_{\varphi} S_{\Phi})\bigg], \\
% ------------
 \langle x^2 y^2 \rangle &=\frac{1}{4}\Bigg[\frac{3 (2 n (n+1)+1) C^2_{\theta}C^2_{\Phi} (C_{\theta } C_{\varphi } S_{\Phi }+S_{\theta } S_{\varphi })^2}{\vartheta _x^2}+\frac{3 (2 m (m+1)+1) C^2_{\varphi } S^2_{\Phi} C^2_{\Phi}}{\vartheta _y^2}\notag\\
 &\phantom{=}+\frac{3 (2 l (l+1)+1) S^2_{\theta } C^2_{\Phi} (C_{\theta } S_{\varphi }-S_{\theta } C_{\varphi } S_{\Phi })^2}{\vartheta _z^2}\Bigg]+\frac{(2 l+1) (2 n+1) C^2_{\Phi} \left(3 S_{\Phi } \left(8 S^2_{\theta } C^2_{\theta}C^2_{\varphi } S_{\Phi }-S_{4\theta } S_{2\varphi }\right)+(3 C_{4 \theta }+1) S^2_{\varphi }\right)}{16 \vartheta _x \vartheta _z}\notag\\
 &\phantom{=}+(1+ 2m) \Bigg[\frac{(2 n+1) \left(4 S^2_{\theta } S^2_{\varphi } S^2_{\Phi}+C^2_{\theta}C^2_{\varphi } (3 C_{4 \Phi }+1)+S_{\theta } C_{\theta } S_{2\varphi } (S_{\Phi }-3 S_{3\Phi })\right)}{16 \vartheta _x \vartheta _y}\notag\\
 &\phantom{=}+\frac{(2 l+1) \left(S^2_{\theta } C^2_{\varphi } (3 C_{4 \Phi }+1)+4 C^2_{\theta}S^2_{\varphi } S^2_{\Phi}-S_{2 \theta} S_{\varphi } C_{\varphi } (S_{\Phi }-3 S_{3\Phi })\right)}{16 \vartheta _y \vartheta _z}\Bigg],
 \\
 \langle x^2 z^2 \rangle &=\frac{1}{4}\Bigg[ \frac{3 (2 n (n+1)+1) C^2_{\theta} C^2_{ \Phi} (S_{\theta} C_{\varphi}-C_{\theta} S_{\varphi} S_{ \Phi})^2}{\vartheta _x^2}+\frac{3 (2 m (m+1)+1) S^2_{ \varphi} S^2_{\Phi} C^2_{ \Phi}}{\vartheta _y^2}\notag\\
 &\phantom{=}+\frac{3 (2 l (l+1)+1) S^2_{\theta} C^2_{ \Phi} (S_{\theta} S_{\varphi} S_{ \Phi}+C_{\theta} C_{\varphi})^2}{\vartheta _z^2} \Bigg]+\frac{(2 l+1) (2 n+1) C^2_{ \Phi} \left(3 S_{ \Phi} \left(2 S^2_{2 \theta} S^2_{ \varphi} S_{ \Phi}+S_{4 \theta} S_{2 \varphi }\right)+(3 C_{4 \theta}+1) C^2_{\varphi }\right)}{16 \vartheta _x \vartheta _z}\notag\\
 &\phantom{=}+(1 + 2  m) \Bigg[\frac{(2 n+1) \left(C^2_{\theta} S^2_{ \varphi} (3 C_{4 \Phi}+1)+4 S^2_{\theta} C^2_{\varphi } S^2_{\Phi}-S_{\theta} C_{\theta} S_{2 \varphi } (S_{ \Phi}-3 S_{3 \Phi})\right)}{16 \vartheta _x \vartheta _y}\notag\\
 &\phantom{=}+\frac{(2 l+1) \left(S^2_{\theta} S^2_{ \varphi} C^4_{\Phi}-4 S_{\theta} S_{\varphi} S_{ \Phi} C^2_{ \Phi} (S_{\theta} S_{\varphi} S_{ \Phi}+C_{\theta} C_{\varphi})+S^2_{\Phi} (S_{\theta} S_{\varphi} S_{ \Phi}+C_{\theta} C_{\varphi})^2\right)}{4 \vartheta _y \vartheta _z} \Bigg],
 \\
 \langle y^2 z^2 \rangle & = \frac{1}{4}\Bigg[\frac{3 (2 n (n+1)+1) (C_{\theta} C_{\varphi} S_{\Phi}+S_{\theta} S_{\varphi})^2 (S_{\theta} C_{\varphi}-C_{\theta} S_{\varphi} S_{\Phi})^2}{\vartheta _x^2}+\frac{3 (2 m (m+1)+1) S^2_{\varphi} C^2_{\varphi} C^4_{\Phi }}{\vartheta _y^2}\notag \\
 &\phantom{=}+\frac{3 (2 l (l+1)+1) (C_{\theta} S_{\varphi}-S_{\theta} C_{\varphi} S_{\Phi})^2 (S_{\theta} S_{\varphi} S_{\Phi}+C_{\theta} C_{\varphi})^2}{\vartheta _z^2} \Bigg]\notag \\
 &\phantom{=}+ (1 + 2 m)\Bigg[\frac{(2 n+1) C^2_{\Phi} \left(3 S_{\Phi} \left(8 C^2_{\theta} S^2_{\varphi} C^2_{\varphi} S_{\Phi}-S_{2 \theta} S_{4  \varphi  }\right)+S^2_{\theta} (3 C_{4 \varphi}+1)\right)}{16 \vartheta _x \vartheta _y}\notag \\
 &\phantom{=}+\frac{(2 l+1) C^2_{\Phi} \left(3 S_{\Phi} \left(2 S^2_{\theta} S^2_{2  \varphi  } S_{\Phi}+S_{2 \theta} S_{4  \varphi  }\right)+C^2_{\theta} (3 C_{4 \varphi}+1)\right)}{16 \vartheta _y \vartheta _z} \Bigg]\notag \\
 &\phantom{=}+\frac{(1 + 2 l) (1 + 2  n) }{2048 \vartheta _x \vartheta _z} \Bigg[-48 S_{4  \theta } S_{4  \varphi  } (S_{3  \Phi }-7 S_{\Phi})-6 C_{4 \theta} (4 (7 C_{4 \varphi}+1) C_{2 \Phi }+3)+24 S^2_{2  \theta } S^2_{2  \varphi  } C_{4 \Phi}\notag \\
 &\phantom{=}+105 (S_{4  \theta } S_{4  \phi }+C_{4 \theta} C_{4  \phi })+105 (C_{4 \theta} C_{4  \phi }-S_{4  \theta } S_{4  \phi })-6 C_{4 \varphi} (4 C_{2 \Phi }+3)-40 C_{2 \Phi }+82 \Bigg]
 \\
 \langle p_x^2 p_y^2 \rangle &=\frac{1}{16} (2 l+1) (2 n+1) C^2_{\Phi} \vartheta _x \vartheta _z \Bigg[3 S_{\Phi} \left(8 S^2_{\theta} C^2_{\theta} C^2_{\varphi} S_{\Phi}-S_{4  \theta } S_{2  \varphi }\right)+(3 C_{4 \theta}+1) S^2_{\varphi}\Bigg]\notag \\
 &\phantom{=}+(1 + 2 m)\Bigg[\frac{1}{16} (2 n+1) \vartheta _x \vartheta _y \left(4 S^2_{\theta} S^2_{\varphi} S^2_{\Phi}+C^2_{\theta} C^2_{\varphi} (3 C_{4 \Phi}+1)+S_{\theta} C_{\theta} S_{2  \varphi } (S_{\Phi}-3 S_{3  \Phi })\right)\notag \\
 &\phantom{=}+\frac{1}{16} (2 l+1) \vartheta _y \vartheta _z \left(S^2_{\theta} C^2_{\varphi} (3 C_{4 \Phi}+1)+4 C^2_{\theta} S^2_{\varphi} S^2_{\Phi}-S_{2 \theta} S_{\varphi} C_{\varphi} (S_{\Phi}-3 S_{3  \Phi })\right) \Bigg]\notag \\
 &\phantom{=}+\frac{1}{4}\Bigg[3 (2 l (l+1)+1) S^2_{\theta} C^2_{\Phi} \vartheta _z^2 (C_{\theta} S_{\varphi}-S_{\theta} C_{\varphi} S_{\Phi})^2+3 (2 m (m+1)+1) C^2_{\varphi} S^2_{\Phi} C^2_{\Phi} \vartheta _y^2\notag \\
 &\phantom{=} +3 (2 n (n+1)+1) C^2_{\theta} C^2_{\Phi} \vartheta _x^2 (C_{\theta} C_{\varphi} S_{\Phi}+S_{\theta} S_{\varphi})^2 \Bigg]
 \\
 \langle p_x^2 p_z^2 \rangle &=\frac{1}{16} (2 l+1) (2 n+1) C^2_{\Phi} \vartheta _x \vartheta _z \Bigg[3 S_{\Phi} \left(2 S^2_{2  \theta } S^2_{\varphi} S_{\Phi}+S_{4  \theta } S_{2  \varphi }\right)+(3 C_{4 \theta}+1) C^2_{\varphi}\Bigg]\notag \\
 &\phantom{=}+(1 + 2 m)\Bigg[\frac{1}{16} (2 n+1) \vartheta _x \vartheta _y \left(C^2_{\theta} S^2_{\varphi} (3 C_{4 \Phi}+1)+4 S^2_{\theta} C^2_{\varphi} S^2_{\Phi}-S_{\theta} C_{\theta} S_{2  \varphi } (S_{\Phi}-3 S_{3  \Phi })\right)\notag \\
 &\phantom{=}+\frac{1}{4} (2 l+1) \vartheta _y \vartheta _z \left(S^2_{\theta} S^2_{\varphi} C^4_{\Phi }-4 S_{\theta} S_{\varphi} S_{\Phi} C^2_{\Phi} (S_{\theta} S_{\varphi} S_{\Phi}+C_{\theta} C_{\varphi})+S^2_{\Phi} (S_{\theta} S_{\varphi} S_{\Phi}+C_{\theta} C_{\varphi})^2\right) \Bigg]\notag \\
 &\phantom{=}+\frac{1}{4} \Bigg[3 (2 l (l+1)+1) S^2_{\theta} C^2_{\Phi} \vartheta _z^2 (S_{\theta} S_{\varphi} S_{\Phi}+C_{\theta} C_{\varphi})^2+3 (2 m (m+1)+1) S^2_{\varphi} S^2_{\Phi} C^2_{\Phi} \vartheta _y^2\notag \\
 &\phantom{=}+3 (2 n (n+1)+1) C^2_{\theta} C^2_{\Phi} \vartheta _x^2 (S_{\theta} C_{\varphi}-C_{\theta} S_{\varphi} S_{\Phi})^2\Bigg]
 \\
 \langle p_y^2 p_z^2 \rangle &= \frac{1}{2048}(2 l+1) (2 n+1) \vartheta _x \vartheta _z \Bigg[-48 S_{4  \theta } S_{4  \varphi  } (S_{3  \Phi }-7 S_{\Phi})-6 C_{4 \theta} (4 (7 C_{4 \varphi}+1) C_{2 \Phi }+3)+24 S^2_{2  \theta } S^2_{2  \varphi  } C_{4 \Phi}\notag \\
 &\phantom{=}+105 (S_{4  \theta } S_{4  \phi }+C_{4 \theta} C_{4  \phi })+105 (C_{4 \theta} C_{4  \phi }-S_{4  \theta } S_{4  \phi })-6 C_{4 \varphi} (4 C_{2 \Phi }+3)-40 C_{2 \Phi }+82\Bigg]\notag\\
 &\phantom{=}+(1 + 2 m)\Bigg[\frac{1}{16} (2 n+1) C^2_{\Phi} \vartheta _x \vartheta _y \left(3 S_{\Phi} \left(8 C^2_{\theta} S^2_{\varphi} C^2_{\varphi} S_{\Phi}-S_{2 \theta} S_{4  \varphi  }\right)+S^2_{\theta} (3 C_{4 \varphi}+1)\right)\notag \\
 &\phantom{=}+\frac{1}{16} (2 l+1) C^2_{\Phi} \vartheta _y \vartheta _z \left(3 S_{\Phi} \left(2 S^2_{\theta} S^2_{2  \varphi  } S_{\Phi}+S_{2 \theta} S_{4  \varphi  }\right)+C^2_{\theta} (3 C_{4 \varphi}+1)\right) \Bigg]\notag \\
 &\phantom{=}+\frac{1}{4}\Bigg[3 (2 l (l+1)+1) \vartheta _z^2 (S_{\theta} S_{\varphi} S_{\Phi}+C_{\theta} C_{\varphi})^2 (C_{\theta} S_{\varphi}-S_{\theta} C_{\varphi} S_{\Phi})^2+3 (2 m (m+1)+1) S^2_{\varphi} C^2_{\varphi} C^4_{\Phi } \vartheta _y^2\notag \\
 &\phantom{=} +3 (2 n (n+1)+1) \vartheta _x^2 (C_{\theta} C_{\varphi} S_{\Phi}+S_{\theta} S_{\varphi})^2 (S_{\theta} C_{\varphi}-C_{\theta} S_{\varphi} S_{\Phi})^2 \Bigg]
 \\
 \langle x^2 p_y^2 \rangle &=\frac{(2 l+1) (2 n+1) \left(S^2_{\theta} C^2_{\Phi} \vartheta _x^2 \vartheta _y (C_{\theta} C_{\varphi} S_{\Phi}+S_{\theta} S_{\varphi})^2+C^2_{\theta} C^2_{\Phi} \vartheta _y \vartheta _z^2 (C_{\theta} S_{\varphi}-S_{\theta} C_{\varphi} S_{\Phi})^2\right)}{4 \vartheta _x \vartheta _y \vartheta _z}\notag\\
 &\phantom{=}+(1 + 2  m) \Bigg[\frac{(2 n+1) \left(S^2_{\Phi} \vartheta _x^2 \vartheta _z (C_{\theta} C_{\varphi} S_{\Phi}+S_{\theta} S_{\varphi})^2+C^2_{\theta} C^2_{\varphi} C^4_{\Phi } \vartheta _y^2 \vartheta _z\right)}{4 \vartheta _x \vartheta _y \vartheta _z}\notag\\
 &\phantom{=}+\frac{(2 l+1) \left(S^2_{\theta} C^2_{\varphi} C^4_{\Phi } \vartheta _x \vartheta _y^2+S^2_{\Phi} \vartheta _x \vartheta _z^2 (C_{\theta} S_{\varphi}-S_{\theta} C_{\varphi} S_{\Phi})^2\right)}{4 \vartheta _x \vartheta _y \vartheta _z} \Bigg]+ \frac{C^2_{\Phi}}{4}\Bigg[ (2 l (l+1)+1) S^2_{\theta} (C_{\theta} S_{\varphi}-S_{\theta} C_{\varphi} S_{\Phi})^2\notag\\
 &\phantom{=}+(2 m (m+1)+1) C^2_{\varphi} S^2_{\Phi}+(2 n (n+1)+1) C^2_{\theta} (C_{\theta} C_{\varphi} S_{\Phi}+S_{\theta} S_{\varphi})^2\Bigg]
 \\
 \langle y^2 p_x^2 \rangle &=\frac{(2 l+1) (2 n+1) \left(C^2_{\theta} C^2_{\Phi} \vartheta _x^2 \vartheta _y (C_{\theta} S_{\varphi}-S_{\theta} C_{\varphi} S_{\Phi})^2+S^2_{\theta} C^2_{\Phi} \vartheta _y \vartheta _z^2 (C_{\theta} C_{\varphi} S_{\Phi}+S_{\theta} S_{\varphi})^2\right)}{4 \vartheta _x \vartheta _y \vartheta _z}\notag\\
 &\phantom{=}+(1 + 2 m)\Bigg[\frac{(2 n+1) \left(C^2_{\theta} C^2_{\varphi} C^4_{\Phi } \vartheta _x^2 \vartheta _z+S^2_{\Phi} \vartheta _y^2 \vartheta _z (C_{\theta} C_{\varphi} S_{\Phi}+S_{\theta} S_{\varphi})^2\right)}{4 \vartheta _x \vartheta _y \vartheta _z}\notag\\
 &\phantom{=}+\frac{(2 l+1) \left(S^2_{\Phi} \vartheta _x \vartheta _y^2 (C_{\theta} S_{\varphi}-S_{\theta} C_{\varphi} S_{\Phi})^2+S^2_{\theta} C^2_{\varphi} C^4_{\Phi } \vartheta _x \vartheta _z^2\right)}{4 \vartheta _x \vartheta _y \vartheta _z} \Bigg]\notag\\
 &\phantom{=}+ \frac{C^2_{\Phi}}{4}\Bigg[(2 l (l+1)+1) S^2_{\theta} (C_{\theta} S_{\varphi}-S_{\theta} C_{\varphi} S_{\Phi})^2+(2 m (m+1)+1) C^2_{\varphi} S^2_{\Phi}+(2 n (n+1)+1) C^2_{\theta} (C_{\theta} C_{\varphi} S_{\Phi}+S_{\theta} S_{\varphi})^2 \Bigg]
 \\
 \langle x^2 p_z^2 \rangle &=\frac{(2 l+1) (2 n+1) \left(S^2_{\theta} C^2_{\Phi} \vartheta _x^2 \vartheta _y (S_{\theta} C_{\varphi}-C_{\theta} S_{\varphi} S_{\Phi})^2+C^2_{\theta} C^2_{\Phi} \vartheta _y \vartheta _z^2 (S_{\theta} S_{\varphi} S_{\Phi}+C_{\theta} C_{\varphi})^2\right)}{4 \vartheta _x \vartheta _y \vartheta _z}\notag\\
 &\phantom{=}+ (1 + 2 m)\Bigg[\frac{(2 n+1) \left(S^2_{\Phi} \vartheta _x^2 \vartheta _z (S_{\theta} C_{\varphi}-C_{\theta} S_{\varphi} S_{\Phi})^2+C^2_{\theta} S^2_{\varphi} C^4_{\Phi } \vartheta _y^2 \vartheta _z\right)}{4 \vartheta _x \vartheta _y \vartheta _z}\notag\\
 &\phantom{=}+\frac{(2 l+1) \left(S^2_{\theta} S^2_{\varphi} C^4_{\Phi } \vartheta _x \vartheta _y^2+S^2_{\Phi} \vartheta _x \vartheta _z^2 (S_{\theta} S_{\varphi} S_{\Phi}+C_{\theta} C_{\varphi})^2\right)}{4 \vartheta _x \vartheta _y \vartheta _z} \Bigg]\notag \\
 &\phantom{=}+\frac{C^2_{\Phi}}{4} \Bigg[(2 l (l+1)+1) S^2_{\theta} (S_{\theta} S_{\varphi} S_{\Phi}+C_{\theta} C_{\varphi})^2+(2 m (m+1)+1) S^2_{\varphi} S^2_{\Phi}+(2 n (n+1)+1) C^2_{\theta} (S_{\theta} C_{\varphi}-C_{\theta} S_{\varphi} S_{\Phi})^2\Bigg]
 \\
 \langle z^2 p_x^2 \rangle &=\frac{(2 l+1) (2 n+1) \left(C^2_{\theta} C^2_{\Phi} \vartheta _x^2 \vartheta _y (S_{\theta} S_{\varphi} S_{\Phi}+C_{\theta} C_{\varphi})^2+S^2_{\theta} C^2_{\Phi} \vartheta _y \vartheta _z^2 (S_{\theta} C_{\varphi}-C_{\theta} S_{\varphi} S_{\Phi})^2\right)}{4 \vartheta _x \vartheta _y \vartheta _z}\notag\\
 &\phantom{=}+(1 + 2 m)\Bigg[\frac{(2 n+1) \left(C^2_{\theta} S^2_{\varphi} C^4_{\Phi } \vartheta _x^2 \vartheta _z+S^2_{\Phi} \vartheta _y^2 \vartheta _z (S_{\theta} C_{\varphi}-C_{\theta} S_{\varphi} S_{\Phi})^2\right)}{4 \vartheta _x \vartheta _y \vartheta _z}\notag\\
 &\phantom{=}+\frac{(2 l+1) \left(S^2_{\Phi} \vartheta _x \vartheta _y^2 (S_{\theta} S_{\varphi} S_{\Phi}+C_{\theta} C_{\varphi})^2+S^2_{\theta} S^2_{\varphi} C^4_{\Phi } \vartheta _x \vartheta _z^2\right)}{4 \vartheta _x \vartheta _y \vartheta _z} \Bigg] \notag\\
 &\phantom{=}+\frac{C^2_{\Phi}}{4} \Bigg[(2 l (l+1)+1) S^2_{\theta} (S_{\theta} S_{\varphi} S_{\Phi}+C_{\theta} C_{\varphi})^2+(2 m (m+1)+1) S^2_{\varphi} S^2_{\Phi}+(2 n (n+1)+1) C^2_{\theta} (S_{\theta} C_{\varphi}-C_{\theta} S_{\varphi} S_{\Phi})^2 \Bigg]
 \\
 \langle y^2 p_z^2 \rangle &= \frac{(2 l (l+1)+1) \vartheta _x \vartheta _y \vartheta _z (S_{\theta} S_{\varphi} S_{\Phi}+C_{\theta} C_{\varphi})^2 (C_{\theta} S_{\varphi}-S_{\theta} C_{\varphi} S_{\Phi})^2+(2 m (m+1)+1) S^2_{\varphi} C^2_{\varphi} C^4_{\Phi } \vartheta _x \vartheta _y \vartheta _z}{4 \vartheta _x \vartheta _y \vartheta _z}\notag\\
 &\phantom{=}+ (1 + 2 m)\Bigg[\frac{(2 n+1) \left(C^2_{\varphi} C^2_{\Phi} \vartheta _x^2 \vartheta _z (S_{\theta} C_{\varphi}-C_{\theta} S_{\varphi} S_{\Phi})^2+S^2_{\varphi} C^2_{\Phi} \vartheta _y^2 \vartheta _z (C_{\theta} C_{\varphi} S_{\Phi}+S_{\theta} S_{\varphi})^2\right)}{4 \vartheta _x \vartheta _y \vartheta _z}\notag\\
 &\phantom{=}+\frac{(2 l+1) \left(S^2_{\varphi} C^2_{\Phi} \vartheta _x \vartheta _y^2 (C_{\theta} S_{\varphi}-S_{\theta} C_{\varphi} S_{\Phi})^2+C^2_{\varphi} C^2_{\Phi} \vartheta _x \vartheta _z^2 (S_{\theta} S_{\varphi} S_{\Phi}+C_{\theta} C_{\varphi})^2\right)}{4 \vartheta _x \vartheta _y \vartheta _z} \Bigg]\notag\\
 &\phantom{=}+\frac{(2 l+1) (2 n+1)}{4 \vartheta _x \vartheta _y \vartheta _z}\Bigg[\vartheta _x^2 \vartheta _y (C_{\theta} S_{\varphi}-S_{\theta} C_{\varphi} S_{\Phi})^2 (S_{\theta} C_{\varphi}-C_{\theta} S_{\varphi} S_{\Phi})^2+\vartheta _y \vartheta _z^2 (C_{\theta} C_{\varphi} S_{\Phi}+S_{\theta} S_{\varphi})^2 (S_{\theta} S_{\varphi} S_{\Phi}+C_{\theta} C_{\varphi})^2 \Bigg]\notag\\
 &\phantom{=}+\frac{1}{4} (2 n (n+1)+1) \Bigg[ (C_{\theta} C_{\varphi} S_{\Phi}+S_{\theta} S_{\varphi})^2 (S_{\theta} C_{\varphi}-C_{\theta} S_{\varphi} S_{\Phi})^2\Bigg]
 \\
 \langle z^2 p_y^2 \rangle &= \frac{(2 l (l+1)+1) \vartheta _x \vartheta _y \vartheta _z (S_{\theta} S_{\varphi} S_{\Phi}+C_{\theta} C_{\varphi})^2 (C_{\theta} S_{\varphi}-S_{\theta} C_{\varphi} S_{\Phi})^2+(2 m (m+1)+1) S^2_{\varphi} C^2_{\varphi} C^4_{\Phi } \vartheta _x \vartheta _y \vartheta _z}{4 \vartheta _x \vartheta _y \vartheta _z}\notag\\
 &\phantom{=}+(1 + 2 m)\Bigg[\frac{(2 n+1) \left(S^2_{\varphi} C^2_{\Phi} \vartheta _x^2 \vartheta _z (C_{\theta} C_{\varphi} S_{\Phi}+S_{\theta} S_{\varphi})^2+C^2_{\varphi} C^2_{\Phi} \vartheta _y^2 \vartheta _z (S_{\theta} C_{\varphi}-C_{\theta} S_{\varphi} S_{\Phi})^2\right)}{4 \vartheta _x \vartheta _y \vartheta _z}\notag\\
 &\phantom{=}+\frac{(2 l+1) \left(C^2_{\varphi} C^2_{\Phi} \vartheta _x \vartheta _y^2 (S_{\theta} S_{\varphi} S_{\Phi}+C_{\theta} C_{\varphi})^2+S^2_{\varphi} C^2_{\Phi} \vartheta _x \vartheta _z^2 (C_{\theta} S_{\varphi}-S_{\theta} C_{\varphi} S_{\Phi})^2\right)}{4 \vartheta _x \vartheta _y \vartheta _z} \Bigg]\notag\\
 &\phantom{=}+\frac{(2 l+1) (2 n+1)}{4 \vartheta _x \vartheta _y \vartheta _z}\Bigg[\vartheta _x^2 \vartheta _y (C_{\theta} C_{\varphi} S_{\Phi}+S_{\theta} S_{\varphi})^2 (S_{\theta} S_{\varphi} S_{\Phi}+C_{\theta} C_{\varphi})^2+\vartheta _y \vartheta _z^2 (C_{\theta} S_{\varphi}-S_{\theta} C_{\varphi} S_{\Phi})^2 (S_{\theta} C_{\varphi}-C_{\theta} S_{\varphi} S_{\Phi})^2 \Bigg]\notag\\
 &\phantom{=}+\frac{1}{4} (2 n (n+1)+1) \Bigg[(C_{\theta} C_{\varphi} S_{\Phi}+S_{\theta} S_{\varphi})^2 (S_{\theta} C_{\varphi}-C_{\theta} S_{\varphi} S_{\Phi})^2 \Bigg].
\end{align}
% \intertext{\newpage} % Force a page break here inside align command

\section*{SECOND Appendix II: Quantum STEERING }
The expressions of quantum steering for the weak coupling regime are detailed below;

\begin{align}
S_{x \rightarrow y} = &-\frac{1}{2 \left(\mu _{\theta }^2+1\right)^2 \left(\mu _{\varphi }^2+1\right) \left(\mu _{\Phi }^2+1\right)^2}\Biggl[\mu _{\theta }^4 \left(l \left(2 m+1\right) \mu _{\Phi }^4-\left(l+m\right) \mu _{\Phi }^2+n \mu _{\varphi }^2 \left(\mu _{\Phi }^2+1\right) \left(2 l+\left(2 m+1\right) \mu _{\Phi }^2+1\right)+2 l m+m\right) \notag \\
&\phantom{=}+2 \mu _{\theta }^3 \mu _{\varphi } \sqrt{\mu _{\Phi }^2+1} \mu _{\Phi } \left(-\left(2 m+1\right) \left(l-n\right) \mu _{\Phi }^2+2 l n+l+n\right)-2 \mu _{\theta } \mu _{\varphi } \mu _{\Phi } \sqrt{\mu _{\Phi }^2+1} \left(\left(2 m+1\right) \left(l-n\right) \mu _{\Phi }^2+2 l n+l+n\right) \notag \\
&\phantom{=}+\mu _{\theta }^2 \biggl(\left(2 m+1\right) \left(l+n\right) \left(\mu _{\varphi }^2+1\right) \mu _{\Phi }^4+\mu _{\Phi }^2 \biggl(2 m \left(l+n\right) \mu _{\varphi }^2+4 l n+l-2 m+n\bigr)+2 m \left(l+n+1\right)-\left(l+n\right) \mu _{\varphi }^2\biggr) \notag \\
&\phantom{=}+l \mu _{\varphi }^2 \left(\mu _{\Phi }^2+1\right) \left(\left(2 m+1\right) \mu _{\Phi }^2+2 n+1\right)+\left(2 m+1\right) n \mu _{\Phi }^4-\mu _{\Phi }^2 \left(m+n\right)+2 m n+m\Biggr],
\end{align}

{\small \begin{align}
S_{y \rightarrow x} =&-\frac{1}{2 \left(\mu _{\theta }^2+1\right)^2 \left(\mu _{\varphi }^2+1\right) \left(\mu _{\Phi }^2+1\right)^2}\Bigg[\mu _{\theta }^4 \biggl(\left(2 l m+m\right) \mu _{\Phi }^4-\left(l+m\right) \mu _{\Phi }^2+\left(2 n+1\right) \mu _{\varphi }^2 \left(\mu _{\Phi }^2+1\right) \left(l+m \mu _{\Phi }^2\right)+2 l m+l\biggr) \notag \\
    &\phantom{=}-4 \mu _{\theta }^3 \mu _{\varphi } \mu _{\Phi } \sqrt{\mu _{\Phi }^2+1} \biggl(m \left(l-n\right) \mu _{\Phi }^2-l \left(n+1\right)\biggr)-4 \mu _{\theta } \mu _{\varphi } \mu _{\Phi } \sqrt{\mu _{\Phi }^2+1} \biggl(m \left(l-n\right) \mu _{\Phi }^2+\left(l+1\right) n\biggr)\notag \\
    &\phantom{=} +\mu _{\theta }^2 \biggl(\mu _{\varphi }^2 \left(\mu _{\Phi }^2+1\right) \left(2 m \left(l+n+1\right) \mu _{\Phi }^2-l-n\right)+2 m \left(l+n+1\right) \mu _{\Phi }^4+\mu _{\Phi }^2 \left(4 l n+l-2 m+n\right)+\left(2 m+1\right) \left(l+n\right)\biggr) \notag \\
     &\phantom{=}+\left(2 l+1\right) \mu _{\varphi }^2 \left(\mu _{\Phi }^2+1\right) \left(m \mu _{\Phi }^2+n\right)+m \mu _{\Phi }^4-m \mu _{\Phi }^2+2 m n \mu _{\Phi }^4+2 m n-n \mu _{\Phi }^2+n \Bigg],\\
S_{x \rightarrow z} =&-\frac{1}{2 \left(\mu _{\theta }^2+1\right)^2\left(\mu _{\varphi }^2+1\right) \left(\mu _{\Phi }^2+1\right)^2} \Bigg[ \mu _{\theta }^4 \left(\mu _{\varphi }^2 \left(\left(2 l m+l\right) \mu _{\Phi }^4-\left(l+m\right) \mu _{\Phi }^2+2 l m+m\right)+n \left(\mu _{\Phi }^2+1\right) \left(2 l+\left(2 m+1\right) \mu _{\Phi }^2+1\right)\right)\notag \\
  &\phantom{=}-2 \mu _{\theta }^3 \mu _{\varphi } \mu _{\Phi } \sqrt{\mu _{\Phi }^2+1} \left(-\left(\left(2 m+1\right) \left(l-n\right) \mu _{\Phi }^2\right)+2 l n+l+n\right)+2 \mu _{\theta } \mu _{\varphi } \sqrt{\mu _{\Phi }^2+1} \mu _{\Phi } \left(\left(2 m+1\right) \left(l-n\right) \mu _{\Phi }^2+2 l n+l+n\right) \notag \\
  &\phantom{=}+\mu _{\theta }^2 \left(\mu _{\varphi }^2 \left(\left(2 m+1\right) \left(l+n\right) \mu _{\Phi }^4+\mu _{\Phi }^2 \left(4 l n+l-2 m+n\right)+2 m \left(l+n+1\right)\right)+\left(l+n\right) \left(\mu _{\Phi }^2+1\right) \left(\left(2 m+1\right) \mu _{\Phi }^2-1\right)\right)\notag \\
  &\phantom{=}+l \left(\mu _{\Phi }^2+1\right) \left(\left(2 m+1\right) \mu _{\Phi }^2+2 n+1\right)+\mu _{\varphi }^2 \left(\mu _{\Phi }^4 \left(2 m n+n\right)-\mu _{\Phi }^2 \left(m+n\right)+2 m n+m\right) \Bigg],\\
S_{z \rightarrow x} =& -\frac{1}{2 \left(\mu _{\theta }^2+1\right){}^2 \left(\mu _{\varphi }^2+1\right) \left(\mu _{\Phi }^2+1\right)^2} \Bigg[\mu _{\theta }^4 \left(\mu _{\varphi }^2 \left(\left(2 l+1\right) m \mu _{\Phi }^4-\left(l+m\right) \mu _{\Phi }^2+2 l m+l\right)+\left(2 n+1\right) \left(\mu _{\Phi }^2+1\right) \left(l+m \mu _{\Phi }^2\right)\right)\notag\\
    &\phantom{=}+4 \mu _{\theta }^3 \mu _{\varphi } \sqrt{\mu _{\Phi }^2+1} \mu _{\Phi } \left(m \left(l-n\right) \mu _{\Phi }^2-l \left(n+1\right)\right)+4 \mu _{\theta } \mu _{\varphi } \sqrt{\mu _{\Phi }^2+1} \mu _{\Phi } \left(m \left(l-n\right) \mu _{\Phi }^2+\left(l+1\right) n\right)\notag\\
    &\phantom{=}+\mu _{\theta }^2 \left(\mu _{\varphi }^2 \left(2 m \left(l+n+1\right) \mu _{\Phi }^4+\mu _{\Phi }^2 \left(4 l n+l-2 m+n\right)+\left(2 m+1\right) \left(l+n\right)\right)\left(\mu _{\Phi }^2+1\right) \left(2 m \left(l+n+1\right) \mu _{\Phi }^2-l-n\right)\right)\notag\\
    &\phantom{=}+\left(2 l+1\right) \left(\mu _{\Phi }^2+1\right) \left(m \mu _{\Phi }^2+n\right)+\mu _{\varphi }^2 \left(m \left(2 n+1\right) \mu _{\Phi }^4-\mu _{\Phi }^2 \left(m+n\right)+2 m n+n\right) \Bigg],\\
S_{y \rightarrow z} =& -\frac{1}{2 \left(\mu _{\theta }^2+1\right){}^2 \left(\mu _{\varphi }^2+1\right){}^2 \left(\mu _{\Phi }^2+1\right)^2}\Bigg[\mu _{\theta }^4 \mu _{\varphi }^2 \left(\mu _{\Phi }^2 \left(4 l m+l+m-2 n\right)-\left(\left(l+n\right) \mu _{\Phi }^4\right)-m-n\right)\notag\\
     &\phantom{=}+2 \mu _{\theta }^3 \mu _{\varphi } \sqrt{\mu _{\Phi }^2+1} \mu _{\Phi } \left(\mu _{\varphi }^2 \left(-2 l m+\left(4 l n+3 l+n\right) \mu _{\Phi }^2+2 l n+l+2 m n+n\right)+2 l m-\left(4 l n+l+3 n\right) \mu _{\Phi }^2-2 l n+l-2 m n-3 n\right)\notag\\ 
     &\phantom{=}-2 \mu _{\theta } \mu _{\varphi } \mu _{\Phi } \sqrt{\mu _{\Phi }^2+1} \left(\mu _{\varphi }^2 \left(2 l m+\left(4 l n+l+3 n\right) \mu _{\Phi }^2+2 l n+l-2 m n+n\right)-2 l m-\left(4 l n+3 l+n\right) \mu _{\Phi }^2-2 l n-3 l+2 m n+n\right)\notag\\
     &\phantom{=}+\mu _{\theta }^2 \mu _{\varphi }^2 \left(\mu _{\Phi }^2 \left(l \left(4 m+24 n+11\right)+4 m n+2 m+11 n\right)+2 \left(l \left(12 n+5\right)+5 n\right) \mu _{\Phi }^4+4 l n+l-2 m+n\right)\notag\\
     &\phantom{=}+\left(2 n+1\right) \left(l \mu _{\Phi }^2 \left(\mu _{\Phi }^2+1\right)+\mu _{\theta }^4 \mu _{\varphi }^4 \left(\mu _{\Phi }^2+1\right) \left(l \mu _{\Phi }^2+m\right)\right)+\left(2 l+1\right) \left(\mu _{\varphi }^4 \left(\mu _{\Phi }^2+1\right) \left(m+n \mu _{\Phi }^2\right)+n \mu _{\theta }^4 \mu _{\Phi }^2 \left(\mu _{\Phi }^2+1\right)\right)\notag\\
     &\phantom{=}+\left(2 m+1\right) \left(l \left(\mu _{\Phi }^2+1\right)+\mu _{\theta }^2 \left(l+n\right) \left(\mu _{\Phi }^2+1\right)+n \mu _{\theta }^4 \left(\mu _{\Phi }^2+1\right)\right)+\mu _{\theta }^2 \mu _{\varphi }^4 \left(\mu _{\Phi }^2+1\right) \left(2 m \left(l+n+1\right)-\left(l+n\right) \mu _{\Phi }^2\right)\notag\\
     &\phantom{=}-\mu _{\varphi }^2 \left(\mu _{\Phi }^2 \left(2 l-4 m n-m-n\right)+\left(l+n\right) \mu _{\Phi }^4+l+m\right)-\mu _{\theta }^2 \left(l+n\right) \mu _{\Phi }^2 \left(\mu _{\Phi }^2+1\right)\Bigg],
\end{align}}

{\small\begin{align}
    S_{z \rightarrow y} =& -\frac{1}{2 \left(\mu _{\theta }^2+1\right){}^2 \left(\mu _{\varphi }^2+1\right){}^2 \left(\mu _{\Phi }^2+1\right){}^2} \Bigg[ 2 l \mu _{\theta } \mu _{\varphi } \sqrt{\mu _{\Phi }^2+1} \mu _{\Phi }+4 l m \mu _{\theta } \mu _{\varphi } \sqrt{\mu _{\Phi }^2+1} \mu _{\Phi }-2 l \mu _{\theta } \mu _{\varphi }^3 \mu _{\Phi } \sqrt{\mu _{\Phi }^2+1} \left(2 m+2 n+3\right)\notag\\
    &+\mu _{\theta }^4 \mu _{\varphi }^2 \left(\mu _{\Phi }^2 \left(4 l m+l+m-2 n\right)-\left(\left(l+n\right) \mu _{\Phi }^4\right)-m-n\right)+4 l n \mu _{\theta } \mu _{\varphi } \sqrt{\mu _{\Phi }^2+1} \mu _{\Phi }-4 m n \mu _{\theta } \mu _{\varphi } \mu _{\Phi } \sqrt{\mu _{\Phi }^2+1}+2 n \mu _{\theta } \mu _{\varphi } \sqrt{\mu _{\Phi }^2+1} \mu _{\Phi }\notag\\
    &-2 \mu _{\theta }^3 \mu _{\varphi } \mu _{\Phi } \sqrt{\mu _{\Phi }^2+1} \left(\mu _{\varphi }^2 \left(2 l m-\left(\left(4 l n+l+3 n\right) \mu _{\Phi }^2\right)-2 l n+l-2 m n-3 n\right)-2 l m+\left(4 l n+3 l+n\right) \mu _{\Phi }^2+2 l n+l+2 m n+n\right)\notag\\
    &+\mu _{\theta }^2 \mu _{\varphi }^2 \left(\mu _{\Phi }^2 \left(l \left(4 m+24 n+11\right)+4 m n+2 m+11 n\right)+2 \left(l \left(12 n+5\right)+5 n\right) \mu _{\Phi }^4+4 l n+l-2 m+n\right)\notag\\
    &+2 \mu _{\theta } \left(4 l n+l+3 n\right) \mu _{\varphi } \sqrt{\mu _{\Phi }^2+1} \mu _{\Phi }^3-2 \mu _{\theta } \left(l \left(4 n+3\right)+n\right) \mu _{\varphi }^3 \mu _{\Phi }^3 \sqrt{\mu _{\Phi }^2+1}+\left(2 n+1\right) \left(l \mu _{\varphi }^4 \mu _{\Phi }^2 \left(\mu _{\Phi }^2+1\right)+\mu _{\theta }^4 \left(\mu _{\Phi }^2+1\right) \left(l \mu _{\Phi }^2+m\right)\right)\notag \\
    &+\left(2 m+1\right) \left(l \mu _{\varphi }^4 \left(\mu _{\Phi }^2+1\right)+\mu _{\theta }^2 \left(l+n\right) \mu _{\varphi }^4 \left(\mu _{\Phi }^2+1\right)+n \mu _{\theta }^4 \mu _{\varphi }^4 \left(\mu _{\Phi }^2+1\right)+2 n \mu _{\theta } \mu _{\varphi }^3 \sqrt{\mu _{\Phi }^2+1} \mu _{\Phi }\right)\notag\\
    &+\left(2 l+1\right) \left(\left(\mu _{\Phi }^2+1\right) \left(m+n \mu _{\Phi }^2\right)+n \mu _{\theta }^4 \mu _{\varphi }^4 \mu _{\Phi }^2 \left(\mu _{\Phi }^2+1\right)\right)+\mu _{\theta }^2 \left(\mu _{\Phi }^2+1\right) \left(2 m \left(l+n+1\right)-\left(l+n\right) \mu _{\Phi }^2\right)\notag\\
    &-\mu _{\varphi }^2 \left(\mu _{\Phi }^2 \left(2 l-4 m n-m-n\right)+\left(l+n\right) \mu _{\Phi }^4+l+m\right)-\mu _{\theta }^2 \left(l+n\right) \mu _{\varphi }^4 \mu _{\Phi }^2 \left(\mu _{\Phi }^2+1\right) \Bigg].
\end{align}}

\end{document}